\documentclass[sigconf,screen]{acmart}

\makeatletter

\makeatletter
\renewcommand\@formatdoi[1]{\ignorespaces}
\makeatother
\renewcommand\footnotetextcopyrightpermission[1]{}

\usepackage{multirow}
\usepackage{tcolorbox}
\usepackage{balance} 
\usepackage{booktabs} 
\usepackage{amsmath}
\usepackage{graphicx}
\usepackage[linesnumbered,ruled,vlined]{algorithm2e}
\usepackage{epstopdf}
\usepackage{afterpage}
\usepackage{listings}
\definecolor{codegreen}{rgb}{0,0.6,0}
\definecolor{codegray}{rgb}{0.5,0.5,0.5}
\definecolor{codepurple}{rgb}{0.58,0,0.82}
\definecolor{backcolour}{rgb}{0.95,0.95,0.92}
\definecolor{myxcodecolour}{rgb}{0.682,0.059,0.588}
\usepackage{listings}
\lstdefinelanguage{GSQL}{
    keywords={CREATE, QUERY, FROM, SELECT, WHERE, ACCUM, PRINT, SumAccum, VERTEX, ALTER, EMBEDDING, ATTRIBUTE, SPACE, PRIMARY, KEY, LOAD, VALUES, VectorSearch,ORDER, BY, LIMIT, ADD },
    keywordstyle=\color{codepurple}\bfseries,
    morecomment=[l]{--},
    morecomment=[l]{//},
    morecomment=[s]{/*}{*/},
    commentstyle=\color{codegreen}\itshape,
    morestring=[b]",
    sensitive=true,
    stringstyle=\color{red},
}

\newcommand{\myline}[1]{{\medskip\noindent\textbf{#1.}}}

\newcommand{\sys}{GraphLake}

\usepackage{url}
\usepackage{enumerate}
\usepackage[labelfont=bf]{caption}
\usepackage{xcolor}
\usepackage{color, colortbl}
\usepackage{graphicx}
\usepackage{subcaption}

\usepackage{pifont}

\lstdefinestyle{mystyle}{
    backgroundcolor=\color{backcolour},   
    commentstyle=\color{codegreen},
    keywordstyle=\color{magenta},
    numberstyle=\tiny\color{codegray},
    stringstyle=\color{codepurple},
    basicstyle=\ttfamily\small,
    breakatwhitespace=false,         
    breaklines=true,                 
    captionpos=b,                    
    keepspaces=true,                 
    numbersep=5pt,                  
    showspaces=false,                
    showstringspaces=false,
    showtabs=false,
    otherkeywords = {<<, >>, WITH}, 
    tabsize=2
}

\begin{document}
\pagestyle{plain} 

\title{GraphLake: A Purpose-Built Graph Compute Engine for \\Lakehouse}

 \author{Shige Liu}
 \affiliation{%
   \institution{Purdue University}
   \country{USA}
 }
 \email{liu3529@purdue.edu}

 \author{Songting Chen}
 \affiliation{%
   \institution{TigerGraph}
   \country{USA}
 }
 \email{songting.chen@tigergraph.com}

 \author{Chengjie Qin}
 \affiliation{%
   \institution{TigerGraph}
   \country{USA}
 }
 \email{chengjie.qin@tigergraph.com}

 \author{Mingxi Wu}
 \affiliation{%
   \institution{TigerGraph}
   \country{USA}
 }
 \email{mingxi.wu@tigergraph.com}

 \author{Jianguo Wang}
 \affiliation{%
   \institution{Purdue University}
   \country{USA}
 }
 \email{csjgwang@purdue.edu}

\begin{abstract}
In this paper, we introduce \sys{}, a purpose-built graph compute engine for Lakehouse. \sys{} is built on top of the commercial graph database TigerGraph. It maps Lakehouse tables to vertex and edge types in a labeled property graph and supports graph analytics over Lakehouse tables using GSQL. To minimize startup time, it loads only the graph topology. Furthermore, it introduces a series of techniques to ensure query efficiency over Lakehouse tables, including a graph-aware caching mechanism and two Lakehouse-optimized parallel primitives. Extensive experiments demonstrate that \sys{} significantly outperforms PuppyGraph, the current state-of-the-art graph compute engine for Lakehouse, by achieving both lower startup and query time.


\end{abstract}

\maketitle

\section{Introduction}\label{sec:intro}


Over the last decade, data management architectures have undergone a significant paradigm shift driven by the growth of cloud storage and the demand for large-scale analytics. Traditional data warehouses provide high performance and robust transactional guarantees; however, they typically rely on proprietary formats that risk vendor lock-in and use tightly coupled architectures that are costly to scale. In contrast, \textit{data lakes}~\cite{RamakrishnanSDK17,HaiKQJ23} emerged as a low-cost alternative by storing data in cloud object storage using open file formats (e.g., Apache Parquet~\cite{Parquet} and Apache ORC~\cite{ORC}) that are not specific to a particular system. Data lakes support elastic scaling through the disaggregation of compute and storage, but they have provided limited support for transactions, schema management, and performance optimization, which can complicate the development of reliable and efficient analytics pipelines.

The \textit{Data Lakehouse} architecture~\cite{LevandoskiCDDEH24,Zaharia0XA21,ArmbrustDPXZ0YM20,BehmPAACDGHJKLL22} has recently emerged to bridge this gap by combining the openness and scalability of data lakes with the reliability and performance features of data warehouses. Central to this architecture is the use of open table formats. Examples include Apache Iceberg~\cite{IcebergBook24,Iceberg}, Delta Lake~\cite{ArmbrustDPXZ0YM20,DeltaLake}, and Apache Hudi~\cite{Hudi}. 
A Lakehouse stores data in cloud object storage using these open table formats (with open file formats such as Apache Parquet~\cite{Parquet} underneath), while adding warehouse-style capabilities including ACID transactions, schema evolution, time travel, and metadata management. This design has gained rapid adoption in industry because it enables multiple processing engines, such as SQL analytics, machine learning, and streaming systems, to operate over the same data without redundant ETL or duplication. As organizations increasingly demand unified platforms for AI and analytics at cloud scale, the Lakehouse has become a foundational architecture in modern data ecosystems by providing a table abstraction.

As a result, there is a trend toward building data systems (mostly OLAP engines) for the Lakehouse, such as Databricks Photon~\cite{BehmPAACDGHJKLL22}, Google BigLake~\cite{LevandoskiCDDEH24}, Microsoft Fabric~\cite{MicrosoftFabric}, and Apache Presto~\cite{Presto23}. Moreover, Amazon Redshift~\cite{RedshiftLakehouse}, Snowflake~\cite{SnowflakeLakehouse}, SingleStore~\cite{SingleStoreLakehouse24}, and DuckDB~\cite{DuckLake25} have started to support Lakehouse architectures.

In this work, we focus on \textit{building a graph analytical (OLAP) engine for the Lakehouse}. In particular, we focus on labeled property graphs (LPGs)~\cite{KondylakisDLYBEFPTTY25,AnglesBD0GHLLMM23,Tian22} due to their wide adoption in real-world applications such as fraud detection, entity resolution, cybersecurity, and customer 360.  In these property graphs, data is modeled as a network of labeled vertices and edges with properties. Examples of property graph systems include TigerGraph~\cite{tigergraph}, Neo4j~\cite{Neo4j}, Amazon Neptune~\cite{Neptune}, NebulaGraph~\cite{NebulaGraph}, {K{\`{U}}ZU~\cite{Kuzu2023}, Google Spanner Graph~\cite{SpannerGraph24}, and IBM Db2 Graph~\cite{Db2Graph2020}.



\begin{figure}[tbp]
\centering
\includegraphics[width=0.47\textwidth]{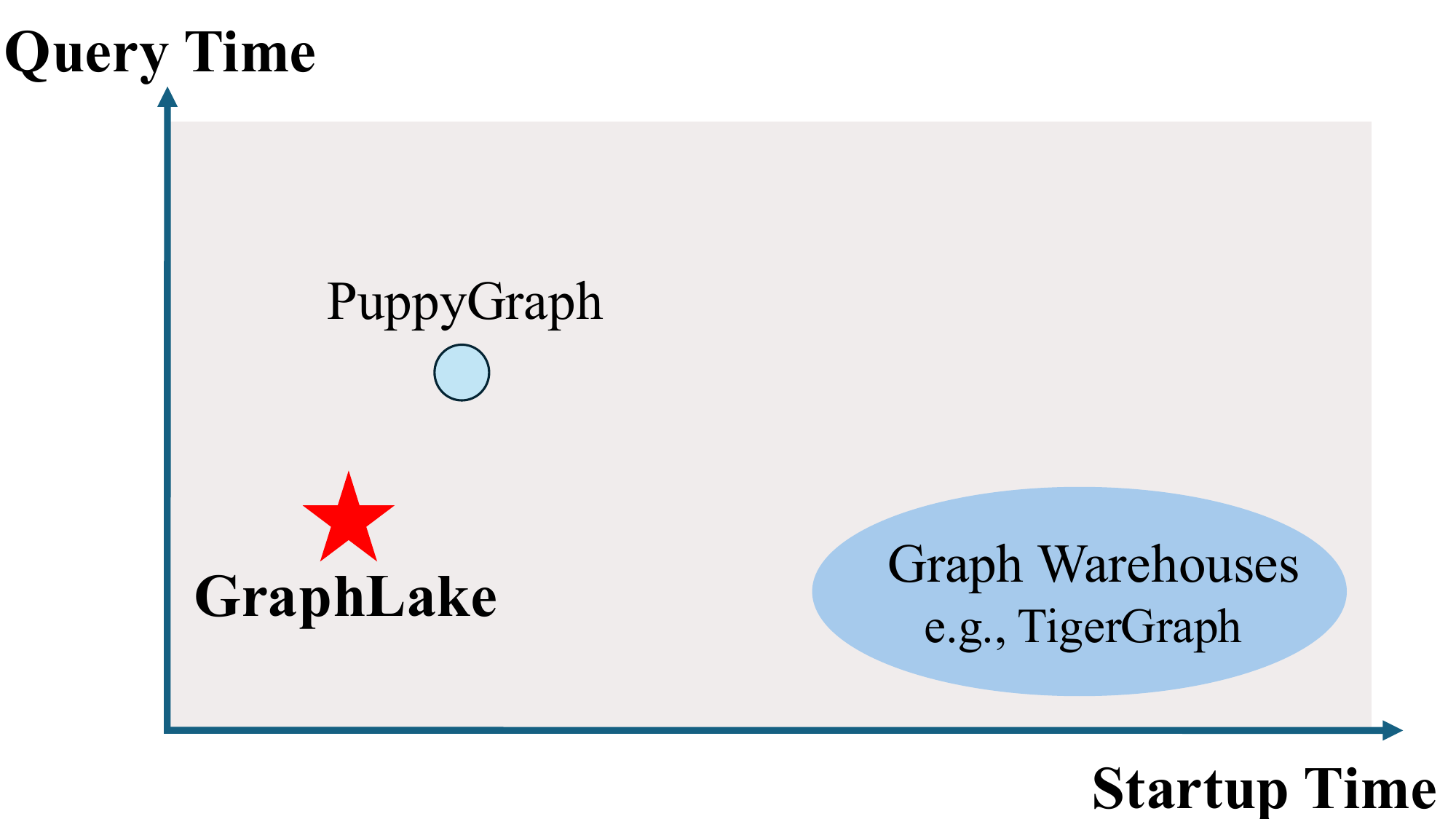}
\caption{Balance of Startup Time and Query Time}\label{fig:tradeoff}
\end{figure}

\myline{Goals}
There are two primary goals in building data systems for the Lakehouse~\cite{BehmPAACDGHJKLL22,LevandoskiCDDEH24,Zaharia0XA21,ArmbrustDPXZ0YM20}: (1) \textit{low startup time} and (2) \textit{low query time}. Low startup time ensures that a system can rapidly provision compute resources and begin processing queries without long cold-start delays. This is especially critical in cloud environment where compute resource is serverless or elastic to deal with dynamic query workload. Low query time, in contrast, focuses on minimizing the runtime execution latency, which is essential for performance-sensitive applications.

\myline{Existing Approaches and Limitations}
Existing approaches (Figure~\ref{fig:tradeoff}) that perform graph  analytics over Lakehouse tables  \textit{either incur high startup time or high query time}. A straightforward solution is to extract data from the Lakehouse and load it into a native graph data management system, such as TigerGraph~\cite{tigergraph}, Neo4j~\cite{Neo4j}, or Amazon Neptune~\cite{Neptune}, and then perform graph analytics within that system. While utilizing an external graph database may offer high performance, doing so forces a departure from the Lakehouse design philosophy of decoupled compute and shared storage. By migrating data into a separate silo, the architecture incurs significant overhead from redundant ETL pipelines and inflated storage costs. This reliance on data duplication not only creates synchronization risks but fundamentally undermines the Lakehouse value proposition: enabling diverse analytics engines to operate on a single, authoritative data source without the burden of maintaining multiple physical copies.

The second approach is in-situ graph analytics, which treats Lakehouse tables as a tiered or remote storage layer and loads data into a stateless compute engine on demand. Conceptually, this resembles a disk-based system, except that the underlying storage resides in the Lakehouse rather than on local disks. To the best of our knowledge, only PuppyGraph~\cite{PuppyGraph} falls into this category that supports Lakehouse tables. This approach benefits from low startup time because it reduces upfront data loading and transformation. However, its query performance is poor due to extensive data movement between storage and compute during query execution. In addition, PuppyGraph is not open source, and many aspects of its system design and optimization techniques are undocumented.

\myline{Challenges}
Building a graph analytics system for the Lakehouse that simultaneously achieves low startup time and low query time is non-trivial. Although prior work on building relational OLAP systems for Lakehouse (e.g., Photon~\cite{BehmPAACDGHJKLL22}, BigLake~\cite{LevandoskiCDDEH24}, and Presto~\cite{Presto23}) provides valuable insights, their techniques cannot be directly applied to graph systems due to fundamental differences between relational and graph data systems. 

Furthermore, distributed graph processing introduces complex sharding and shuffling strategies (e.g., vertex-based or edge-based sharding), which are more challenging than those in relational analytics. 

Lastly, Lakehouse adopts a table-based data model that fundamentally differs from the graph data model. Executing graph computations directly over tables incurs expensive join operations, highlighting the need for new system designs.

\myline{Overview of \sys{}} 
In this paper, we introduce \sys{}, a purpose-built graph compute engine for analytics over Lakehouse tables. It is built on top of TigerGraph~\cite{tigergraph}, a native commercial graph database. It maps Lakehouse tables to vertex and edge types in a graph and seamlessly supports graph analytics with GSQL~\cite{GSQL}, the graph query language of TigerGraph.

To minimize startup time, we propose a new approach in Section~\ref{sec:loading} that decouples the \textit{graph topology} from the graph element properties and loads only the topology during startup. We introduce a new data structure, the \textit{edge list}, to represent the graph topology. This data structure supports fast building and serves as the basis for efficient graph analytics. We also introduce transformed vertex IDs to support efficient attribute retrieval.

To ensure query efficiency, we propose a graph-aware caching mechanism (Section~\ref{sec:cache}) to efficiently access graph element properties in Lakehouse tables. We also introduce two Lakehouse-optimized primitives (Section~\ref{sec:processing}), which leverage edge lists and graph-aware cache units to efficiently support graph analytics over Lakehouse tables, under the compute framework inherited from TigerGraph.


\myline{Experimental Overview}
Experiments on LDBC benchmarks~\cite{ldbc-snb,graphanalytics} show that \sys{} outperforms PuppyGraph with significantly lower startup time and query time. Specifically, \sys{} achieves up to \textbf{26.3}$\times$ faster startup and up to \textbf{60.3}$\times$ faster query time compared to PuppyGraph.

\myline{Contributions} 
This paper is the first work that systematically explores how to build a lightweight and efficient graph compute engine for a Lakehouse. We identify key challenges in this setting and propose a suite of novel techniques to address them. We design and implement \sys{}, a new graph compute engine integrated with TigerGraph that enables efficient graph analytics directly over Lakehouse tables, using Apache Iceberg as a representative Lakehouse table format. We believe that the techniques introduced in this work are broadly applicable to other graph data management systems when operating over Lakehouse tables.

\section{Background}\label{sec:background}

\subsection{Lakehouse Architecture}\label{sec:backgroundlakehouse}

Lakehouse systems are data management platforms built upon low-cost cloud object stores (aka data lakes). They unify data lakes and data warehouses by combining the cost-efficiency of the former with powerful management features of the latter~\cite{Zaharia0XA21}. Modern Lakehouses rely on open table formats (e.g., Apache Iceberg~\cite{Iceberg}, Delta Lake~\cite{ArmbrustDPXZ0YM20}, and Apache Hudi~\cite{Hudi}), which store data in cloud-based data lakes (e.g., Amazon S3, Google Cloud Storage) as immutable files using formats such as Parquet, ORC, or Avro. These systems implement a metadata-driven architecture: updates are executed as logical changes within a metadata layer, leaving the underlying data files untouched. This layer consists of a hierarchy of metadata files managed by a catalog (e.g., Hive Metastore, AWS Glue) that ensures atomic updates, thereby enabling management features such as ACID transactions and schema evolution.

Lakehouses primarily use column-oriented open file formats, such as Parquet and ORC, for tabular data storage. We take Parquet as an illustrative example: A Parquet file includes a schema defining data types and is partitioned horizontally into row groups. Within each row group, values for a specific column are aggregated into a column chunk, which is the fundamental unit for scanning. Each column chunk is individually encoded (e.g., using RLE or dictionary encoding) and stored as a sequence of consecutive pages. While pages support independent decompression and decoding, they often rely on auxiliary data within the chunk (e.g., a dictionary page for dictionary decoding). Finally, file-level metadata, including row counts and byte offsets, is stored in the file footer.

\subsection{Graph Analytics and GSQL in TigerGraph}\label{sec:backgroundgsql}


Graph analytics, also known as graph OLAP, refers to graph analytical workloads that involve (1) computing aggregated measures over vertex and edge properties~\cite{grapholap}, and (2) analyzing the structure and connectivity of graphs, typically via iterative graph algorithms~\cite{bonifati2025roadmap,Tian22}. Graph analytics has widespread applications in social networks, recommendation, security, and fraud detection, where extracting insights requires understanding both the global structure and aggregated properties. For example, graph algorithms such as PageRank identify influential users, while aggregate queries (e.g., calculating the average number of comments each person creates) reveal high-level trends across large subgraphs. Compared to graph OLTP, which typically targets interactive graph queries, graph analytics is characterized by long-running and computationally intensive tasks, often traversing the majority of the graph.



TigerGraph~\cite{tigergraph} is a distributed, native graph database optimized for large-scale graph analytics. It supports cluster-based deployment where data is automatically partitioned into \textit{segments}, each of which contains a fixed number of vertices and serves as the fundamental unit of parallel and distributed computation.
TigerGraph represents the graph topology using Compressed Sparse Row (CSR)~\cite{CSRFormat}, in which a source vertex’s outgoing edges are co-located within its assigned segment.
TigerGraph leverages a massively parallel processing architecture for high-performance graph analytics. Two key parallel primitives, \textit{VertexMap} and \textit{EdgeMap}, enable user-defined functions (UDFs) to execute across the vertices and edges. 


TigerGraph~\cite{tigergraph} employs GSQL~\cite{GSQL}, a Turing-complete language designed for graph analytics. A GSQL query is composed of a sequence of query blocks (\texttt{SELECT-FROM-WHERE}), each producing a vertex set variable. These blocks operate on a vertex-centric computational model, where each block processes an input vertex set and traverses incident edges to produce a subsequent vertex set for the next block.
The \texttt{FROM} clause allows these query blocks to reference variables from prior blocks, facilitating complex query composition. Furthermore, GSQL supports set operations such as \texttt{UNI/ON}, \texttt{INTERSECT}, and \texttt{MINUS} to manipulate these vertex set variables.

Accumulators~\cite{GSQL} are another important tool for complex graph analytics in GSQL.
These runtime variables are mutable throughout the query lifecycle, allowing for the parallel update of complex metrics (e.g., PageRank scores or community IDs) during graph traversal. By combining accumulators with control-flow primitives such as WHILE and IF-ELSE, GSQL enables sophisticated property aggregation and structural analysis during traversal.

\section{System Design Overview}\label{sec:design}

\begin{figure}[tbp]
\centering
\includegraphics[width=0.42\textwidth]{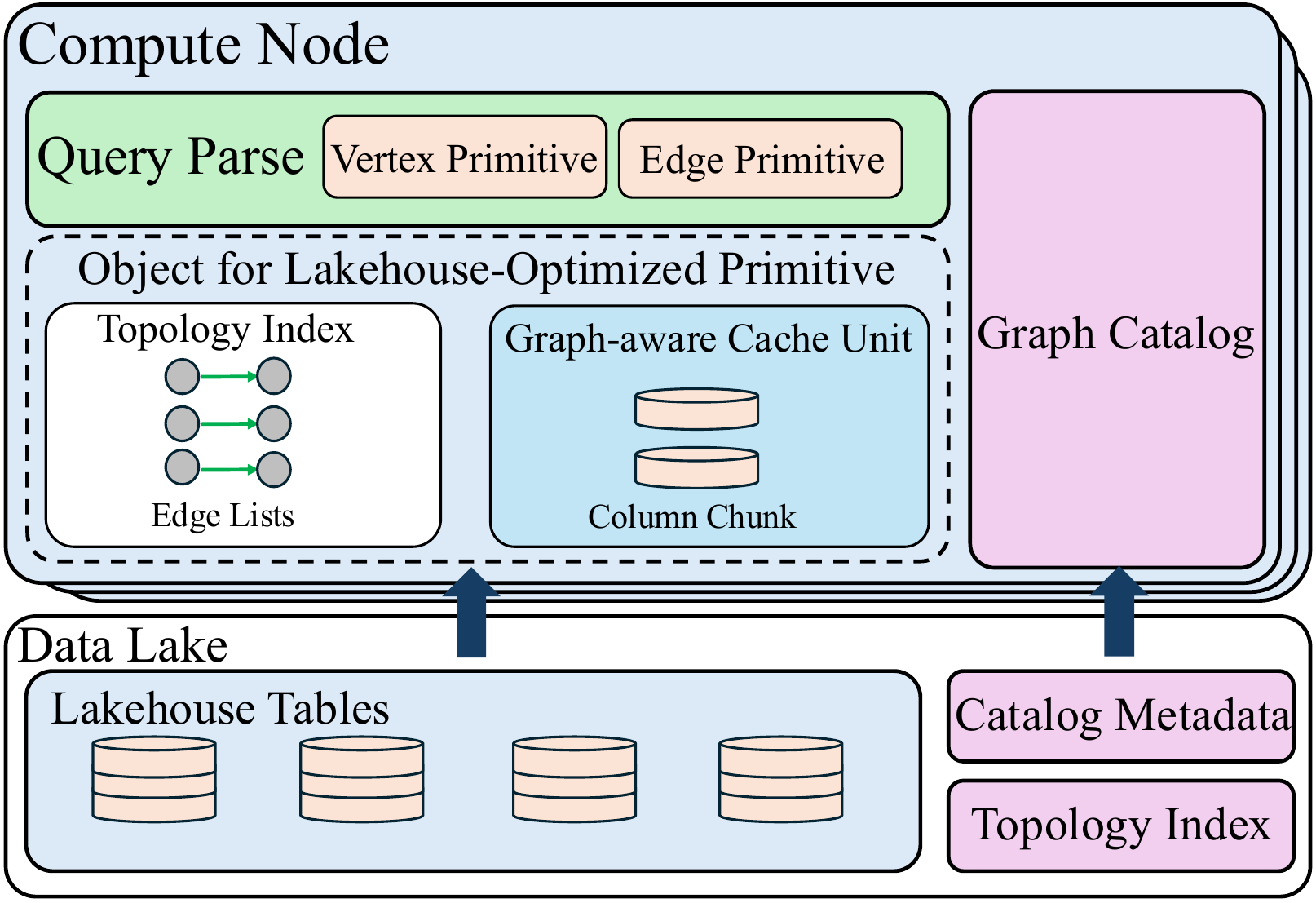}
\caption{System Overview}\label{fig:TigerLake_Overview}
\end{figure}

In this section, we present the system architecture of \sys{}. As illustrated in Figure~\ref{fig:TigerLake_Overview}, \sys{} is a distributed compute engine built upon TigerGraph that enables native graph analytics over Lakehouse tables (e.g., Apache Iceberg). The system executes on compute nodes and interfaces with the underlying storage layer of the Lakehouse. It provides a graph abstraction over Lakehouse tables, allowing users to build graph topology indexes and manage table caches to perform high-performance analytical queries.



\myline{Data Model}
\sys{} is a graph compute engine built on top of TigerGraph, but the \textit{graph representation is derived from Lakehouse-style relational tables}. Following prior work such as GQ-Fast~\cite{lin2016fast}, graphs can be naturally represented using relational tables. Specifically, there are two types of tables: Entity (Vertex) tables and Relationship (Edge) tables. Each entity table contains a primary key column, while each relationship table contains two foreign key columns referencing the primary keys of entity tables. Together, these tables define a graph structure. Tuples in entity tables correspond to vertices, and tuples in relationship tables correspond to edges; columns in both represent properties of vertices and edges. Different types of vertices and edges can be stored in separate tables. Since TigerGraph enforces a strong type system in which each vertex or edge has exactly one type, this design greatly facilitates accurate and efficient graph mapping from Lakehouse tables.


\myline{Compute Framework for Graph Analytics} Building upon TigerGraph, \sys{} leverages the \textit{accumulator-based aggregation paradigm}~\cite{GSQL}, which utilizes polymorphic data containers (accumulators) to store, update, and persist computational states directly on vertices. This framework supports advanced graph analytics by enabling iterative control flow and state composition, allowing users to express both graph-aggregation queries and complex algorithms through declarative state transitions. \sys{} evaluates queries using the Bulk Synchronous Parallel (BSP)~\cite{BSP} model, where execution proceeds through a series of synchronized "supersteps" driven by active vertex sets (or vertex frontiers). To enable elastic scale-out, \sys{} employs file-based partitioning that partitions files across different compute nodes, while vertex accumulators are co-located with their corresponding vertex files in the Lakehouse.





\myline{Graph Catalog} 
To manage the graph defined over Lakehouse tables, we implement a graph catalog module. This module maintains the mapping metadata that links element types (vertex and edge types) to their corresponding Lakehouse tables. It also monitors the underlying Lakehouse tables for changes, such as the addition or deletion of table files, to automatically update the catalog's mapping metadata and related topology indices.


\myline{Topology-only Startup Loading}
To accelerate startup over Lakehouse tables, we propose a new approach that decouples the \textit{graph topology} from the graph element properties (or graph content) and loads only the topology at startup. We introduce a new data structure \textit{edge list} to represent graph topology. This data structure enables fast building and serves as the foundation for efficient graph analytics. We also introduce transformed vertex IDs to support efficient attribute retrieval. This decoupled graph topology design introduces the challenge of accessing property data during query execution, a problem we address in Section~\ref{sec:cache} and Section~\ref{sec:processing}.

\myline{Graph-aware Columnar Caching} 
To enhance query performance, we implement a caching mechanism that utilizes column chunks as the fundamental cache units. We refine these into \textbf{graph-aware cache units}, which minimize decoding overhead by aligning data retrieval with specific graph access patterns. We also propose a graph-aware replacement algorithm to manage the cache allocation under memory and disk constraints. To anticipate and load data before it is accessed during graph traversal, we develop a proactive prefetching mechanism, driven by vertex frontiers and edge list statistics.

\myline{Lakehouse-optimized Parallel Primitives} 
To realize the compute framework over decoupled storage, GraphLake implements two Lakehouse-optimized parallel primitives: \textit{VertexMap} and \textit{EdgeScan}. These primitives are iteratively executed via the BSP model, driven by active vertex sets. Specifically, \textit{VertexMap} applies UDFs to vertices within the active set, whereas \textit{EdgeScan} applies UDFs to neighboring edges by performing a parallel scan over \textit{edge list} structures (Section~\ref{sec:loading}). Both primitives aggregate graph content using graph-aware cache units (Section~\ref{sec:cache}). When deployed on a distributed cluster, \textit{EdgeScan} employs a two-pass method to address the problem that edges and vertices do not co-locate: the first pass identifies remote vertices and the second pass executes the UDFs after the necessary vertex data has been fetched.

\section{Topology-Only Startup Loading}\label{sec:loading}

In this section, we describe how \sys{} achieves fast data loading with low startup overhead while enabling efficient graph analytics over Lakehouse tables.

\myline{High-level Idea}
We propose a new approach that decouples the graph topology from the graph content and loads only the topology at startup. The graph topology consists of the primary key columns from vertex tables and the foreign key columns from edge tables, excluding all vertex and edge properties (or attributes). For example, in Figure~\ref{fig:topologyloading}, we load the \textit{ID} column from the \textit{Person} table and the two foreign key columns from the \textit{Knows} table that reference \textit{Person}, and then construct the graph topology structure.

This design strikes a balance between low startup overhead and high query performance. The graph topology typically occupies only a small fraction of the total table size; for example, the primary key columns of vertex tables in the LDBC\_SNB SF100 dataset~\cite{ldbc-snb} account for only 3\% of the total data volume, as illustrated in Figure~\ref{fig:idsize}. However, the topology is sufficient to evaluate graph analytical queries via traversal. Vertex and edge properties are loaded on demand based on specific query requirements, with details and optimizations discussed in Section~\ref{sec:cache} and Section~\ref{sec:processing}. As a result, even when Lakehouse tables contain extensive attribute sets, \sys{} achieves efficient startup by loading only key columns.

The topology-only startup loading design follows the general philosophy of building data systems over Lakehouse architectures~\cite{Zaharia0XA21}, where the graph topology can be viewed as an auxiliary data structure that enables efficient query processing.


However, the challenge is how to represent the graph topology in a Lakehouse setting, especially in a distributed environment; this challenge is discussed in Section~\ref{sec:optimizedgraphtopo}. We also present optimizations to improve the loading process in Section~\ref{sec:topooptimization} and describe the complete loading workflow in Section~\ref{sec:overalloading}.

\begin{figure}[tbp]
\centering
\includegraphics[width=0.47\textwidth]{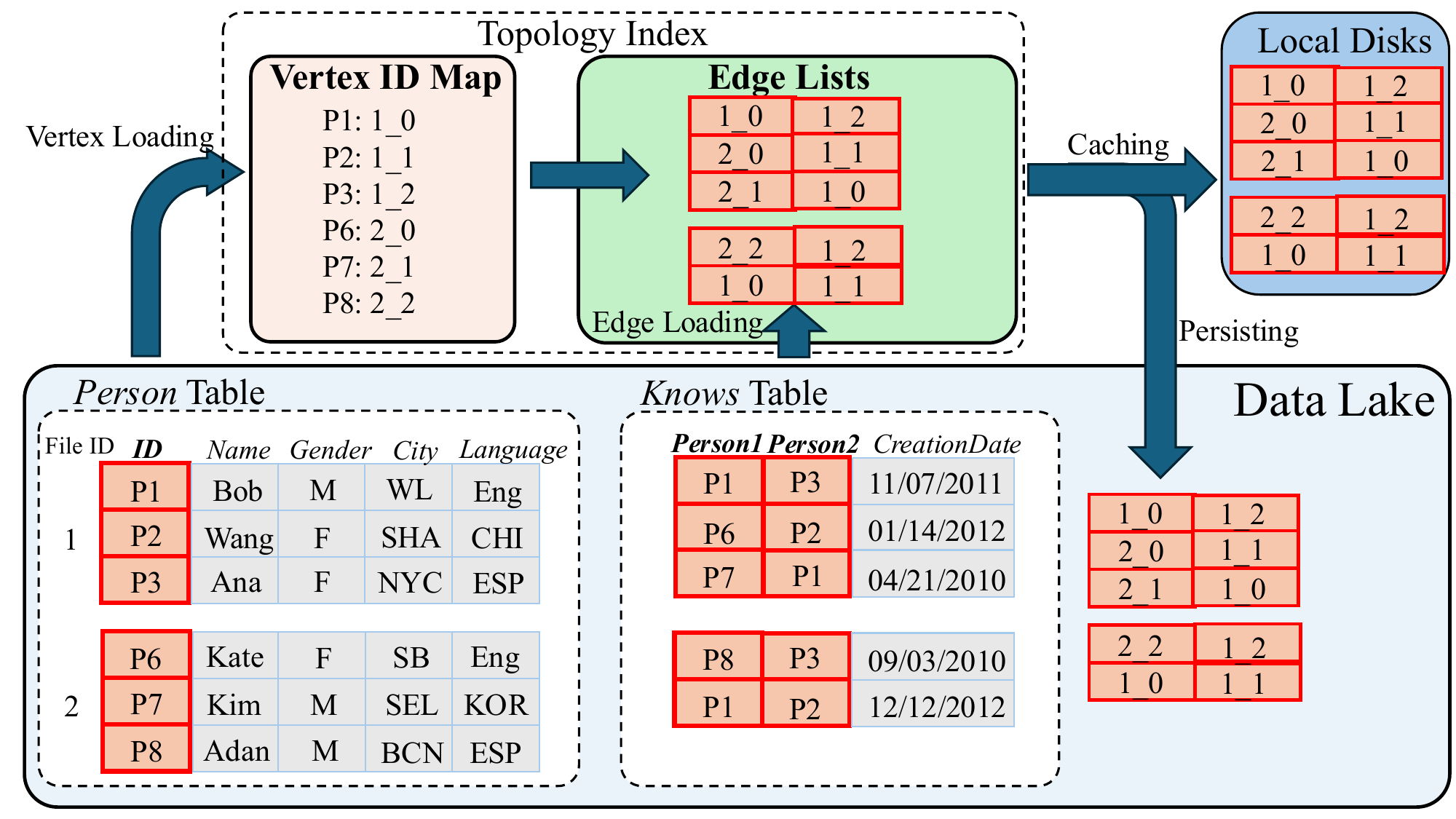}
\caption{Topology-Only Loading}\label{fig:topologyloading}
\end{figure}


\subsection{Lakehouse-Optimized Graph Topology}\label{sec:optimizedgraphtopo}



Next, we describe how to represent the graph topology in the Lakehouse setting.

TigerGraph, like other modern graph databases, internally represents graphs using the CSR structure, a highly compact and vertex-centric layout in which a vertex's outgoing edges are stored contiguously. However, we observe that this representation is not suitable for fast startup in a Lakehouse environment for two primary reasons: (1) it is expensive to build such a data structure, as it requires grouping all edges by source vertex. When deployed in a distributed cluster, edges must be sent over the network to multiple compute nodes, which significantly slows the startup; (2) it is expensive to support updates. For example, when an edge file is deleted, the CSR must identify all source vertices of the deleted edges and update the index structure, shifting large portions of the adjacency array.


In \sys{}, we represent the graph topology using an \textit{edge list}, which is simply a collection of edges derived from the Lakehouse edge table. Specifically, for a given edge table, we load the foreign key columns and construct a set of (source, target) vertex ID pairs as the edge list. As a result, the entries in the edge list preserve the original ordering of the Lakehouse edge table. This facilitates OLAP-style scanning by preserving row-level alignment with edge attributes stored in the Lakehouse, which enables sequential access over edges. For example, when evaluating \textit{Knows} edges in Figure~\ref{fig:topologyloading}, the edge list entries and the corresponding \textit{CreationDate} column are scanned in tandem, ensuring efficient data processing.

Since Lakehouse edge tables are physically partitioned into multiple edge files for scalability and elasticity in distributed environments, we build one edge list per edge file. For example, the \textit{Knows} table in Figure~\ref{fig:topologyloading} has two edge files, with each file corresponding to a unique edge list.

This design provides several advantages. First, it is fast to build, and thus the startup time is low. We only need to download the foreign key columns from edge tables and sequentially scan them to build edge lists. The per-file design also enables efficient parallel construction, as edge lists for different files can be built in parallel.

Second, it is easy to update. Since edge lists are built per edge file, this design enables incremental updates of edge lists. For example, when new edge files are added to the edge table, we only need to build edge lists for the new files without touching other edge lists.


Third, it is more suitable for graph OLAP queries, which are the focus of this paper and tend to evaluate a large number of edges. The edge list design naturally fits an edge-centric processing approach, which iterates over each edge list entry to evaluate the corresponding edge. Compared to the vertex-centric approach based on CSR, it has better CPU cache efficiency. Our experiments in Section~\ref{sec:additional_exp} show that the edge list design outperforms CSR when edge selectivity exceeds 10\%.

Lastly, it is easy to be distributed. We take a file-based partitioning method as mentioned earlier. Each compute node only builds edge lists for its own edge files to enable distributed loading. Edge lists, along with the underlying edge files, are partitioned to enable distributed queries, which will be discussed in Section~\ref{sec:processingdistributed}.

The next challenge is how to efficiently retrieve attributes for vertices and edges. Retrieving edge attributes is straightforward as the edge list maintains row-level alignment with the underlying edge table. However, retrieving vertex attributes is more challenging because a Lakehouse vertex table is randomly partitioned into multiple vertex files. Given a vertex ID, one would need to scan all vertex files to locate the corresponding record, since no indexing structures (e.g., B-trees) are built on top of the Lakehouse vertex table.

To address this challenge, we replace raw vertex IDs in the edge list with \textbf{transformed vertex IDs} designed to optimize attribute lookups. Each transformed ID is globally unique and encodes a file ID combined with the vertex's local row index. Specifically, we represent each transformed ID as a 64-bit integer: the upper 32 bits store the unique file ID, while the lower 32 bits store the row index.
This transformed ID allows efficient point lookup of vertex attributes by directly locating the file ID and row index.

We use a hash map, referred to as the Vertex ID Mapping (Vertex IDM), to store the mapping between raw vertex IDs and their transformed IDs. During edge list building, we leverage the Vertex IDM to translate raw vertex IDs into transformed IDs. For example, in Figure~\ref{fig:topologyloading}, the first edge in the \textit{Knows} table references raw IDs P1 and P3; these are translated into 1\_0 (representing file ID 1, row index 0) and 1\_2 (file ID 1, row index 2) within the edge list, as specified by the Vertex IDM.


In a distributed environment, we replicate the Vertex IDM across all compute nodes because the latency of converting raw vertex IDs to transformed IDs is a critical bottleneck for edge list building. Moreover, since the number of vertices is typically an order of magnitude smaller than the number of edges, the size of the Vertex IDM is usually small, fitting easily within the memory of each compute node.

\begin{figure}[tbp]
\centering
\includegraphics[width=0.35\textwidth]{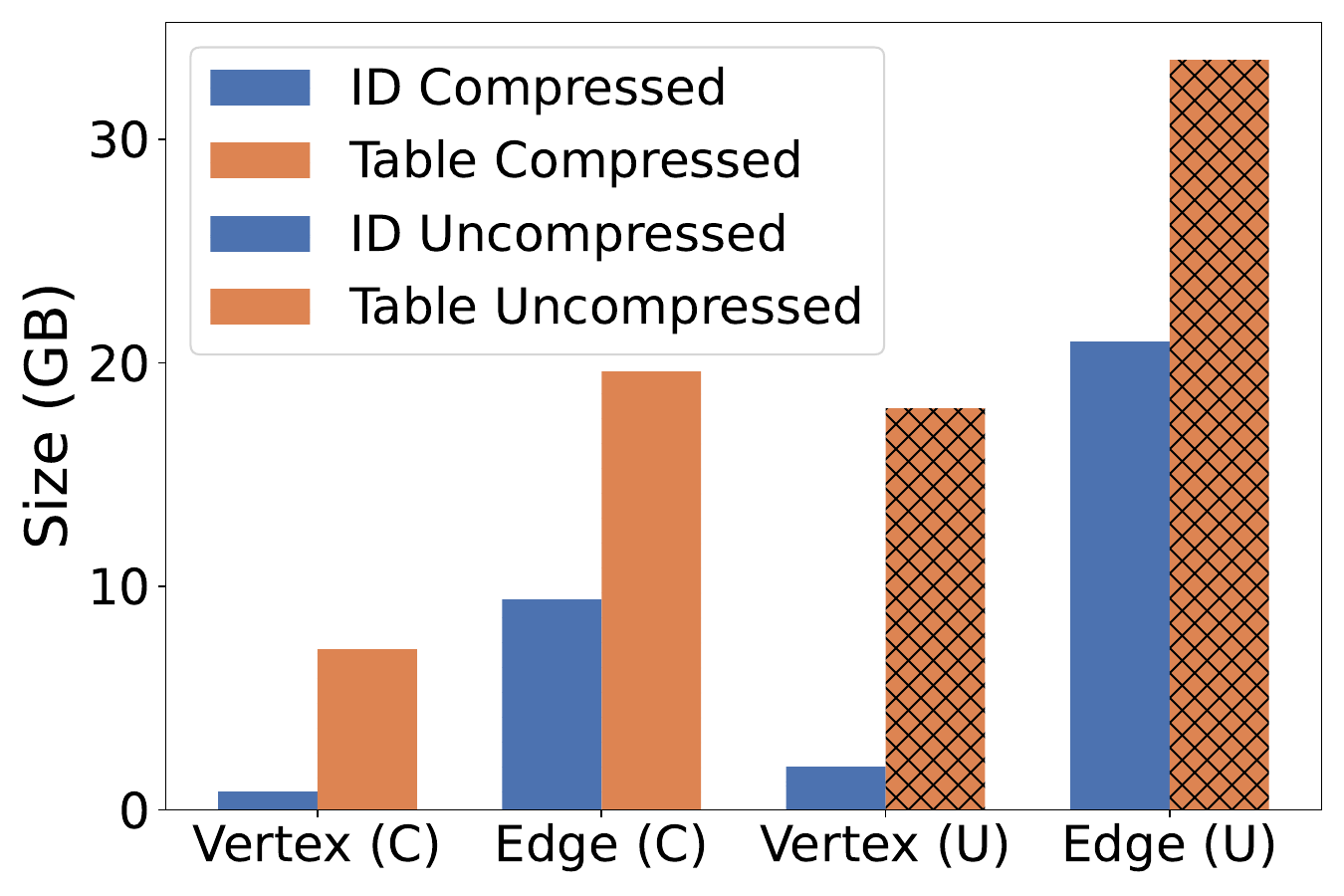}
\caption{ID Column Size vs. Table Size in LDBC\_SNB SF100}\label{fig:idsize}
\end{figure}

\subsection{Optimizations}\label{sec:topooptimization}

We further implement optimizations to improve startup loading.

\myline{Topology Materialization} 
Once the graph topology is built, it is persisted as binary files on the data lake. If compute nodes restart, \sys{} directly loads the pre-stored graph topology without rebuilding it from scratch, which significantly accelerates the startup loading process.

\myline{Pipelining} 
Building the Vertex IDM and edge lists requires downloading key columns from the data lake, which involves multiple steps. We first issue an HTTP request to retrieve the footer's metadata length, followed by a second request for the footer metadata itself; finally, we issue a third request to fetch the specific column chunks required to construct the IDM and edge lists. Although we use multiple threads to process several files concurrently, the high latency of HTTP requests can stall subsequent computation and leave CPUs idle.

To address this, we maintain an I/O thread pool responsible for asynchronous I/O with local disks and the data lake, enabling a pipelined workflow: while I/O threads fetch column chunks or persist edge lists, compute threads concurrently build the Vertex IDM and subsequent edge lists. This task overlapping effectively hides I/O latency and ensures consistent CPU utilization throughout the loading process.

\subsection{The Overall Startup Loading Process}\label{sec:overalloading}

Next, we introduce the overall startup loading process, where multiple threads are leveraged.


During startup loading, \sys{} first sets up connectors to the Lakehouse and the underlying data lake. It identifies all data files for each vertex and edge table, records their file positions in the data lake, and then directly interacts with the data lake.

\myline{Vertex IDM Building} 
Upon identifying all vertex files and assigning unique file IDs, \sys{} initiates the Vertex IDM (ID Mapping) Building. Compute threads scan the downloaded primary key columns of the data files in parallel. For each raw vertex ID, a compute thread calculates its transformed vertex ID based on its row index and the file ID of the data file, then prepares a mapping entry consisting of the raw and transformed IDs. These entries are grouped and inserted into hashmaps in batches to minimize lock contention between compute threads.

\myline{Edge List Building} 
Following the Vertex IDM building, \sys{} proceeds to build the edge lists. These edge lists are built concurrently: each compute thread works on one edge file, scanning its foreign key columns and translating raw vertex IDs into transformed IDs based on the Vertex IDM. Since each thread operates on an independent edge file, the primary building path is lock-free and efficient.

Raw vertex IDs within foreign key columns typically correspond to entries in the Vertex IDM. In certain edge cases, however, we may encounter dangling raw IDs that do not belong to any vertex. To handle these, \sys{} reserves a special file ID (e.g., file ID 0) managed by a global atomic counter. When compute nodes encounter a dangling raw ID, they request a row index from the counter to generate a transformed ID for it. This mechanism ensures that \sys{} maintains complete topology coverage, even when vertices are defined implicitly within edge tables.

Upon completion of edge list building, the Vertex IDM is deallocated to free memory. At runtime, the engine decodes the transformed vertex IDs to perform high-speed lookups of vertex attributes via direct file-path and record-offset addressing.



\section{Graph-Aware Columnar Caching}\label{sec:cache}

As discussed in Section~\ref{sec:loading}, only the graph topology is loaded during the startup process, and graph element properties (both vertex and edge attributes) are loaded on demand during query processing. Thus, it is important to implement an efficient caching mechanism to mitigate I/O latency and maintain high query performance.



Though caching strategies have been studied extensively in existing disk-based graph databases, including~\cite{graphchi, lsmgraph, railwaygraph, FlashGraph}, their caching techniques cannot be applied to graph analytics over Lakehouse. These systems typically utilize page-based caching units, whereas \sys{} relies on columnar chunking in open format files (as explained below). Moreover, existing disk-based graph databases are based on a vertex-centric storage engine that groups all edges of the same vertex together, whereas \sys{} operates on edge lists and underlying columnar files (i.e., an edge-centric design).


In \sys{}, we adopt column chunks as the caching unit, following prior work on relational databases over Lakehouse architectures~\cite{RedshiftLakehouse, DuckLake25, SingleStoreLakehouse24}. As mentioned in Section~\ref{sec:backgroundlakehouse}, each file in Lakehouse tables is horizontally divided into multiple row groups, in which the values of the same column are grouped into column chunks. Thus, retrieving attributes based on column chunks prunes irrelevant columns. Moreover, the size of the column chunks is optimal for efficient network transfer.


However, existing caching techniques in relational databases over Lakehouse~\cite{RedshiftLakehouse, DuckLake25, SingleStoreLakehouse24} cannot address the unique challenges of graph workloads: (1) how to optimize attribute access to accommodate graph access patterns, which is addressed in Section~\ref{sec:cacheunit}; (2) how to manage cache eviction under memory or disk budget constraints, which is addressed in Section~\ref{sec:cacheallocation}; (3) how to prefetch cache units anticipated for future graph traversal, which is addressed in Section~\ref{sec:cacheprefetch}.




\begin{figure}[tbp]
\centering
\includegraphics[width=0.47\textwidth]{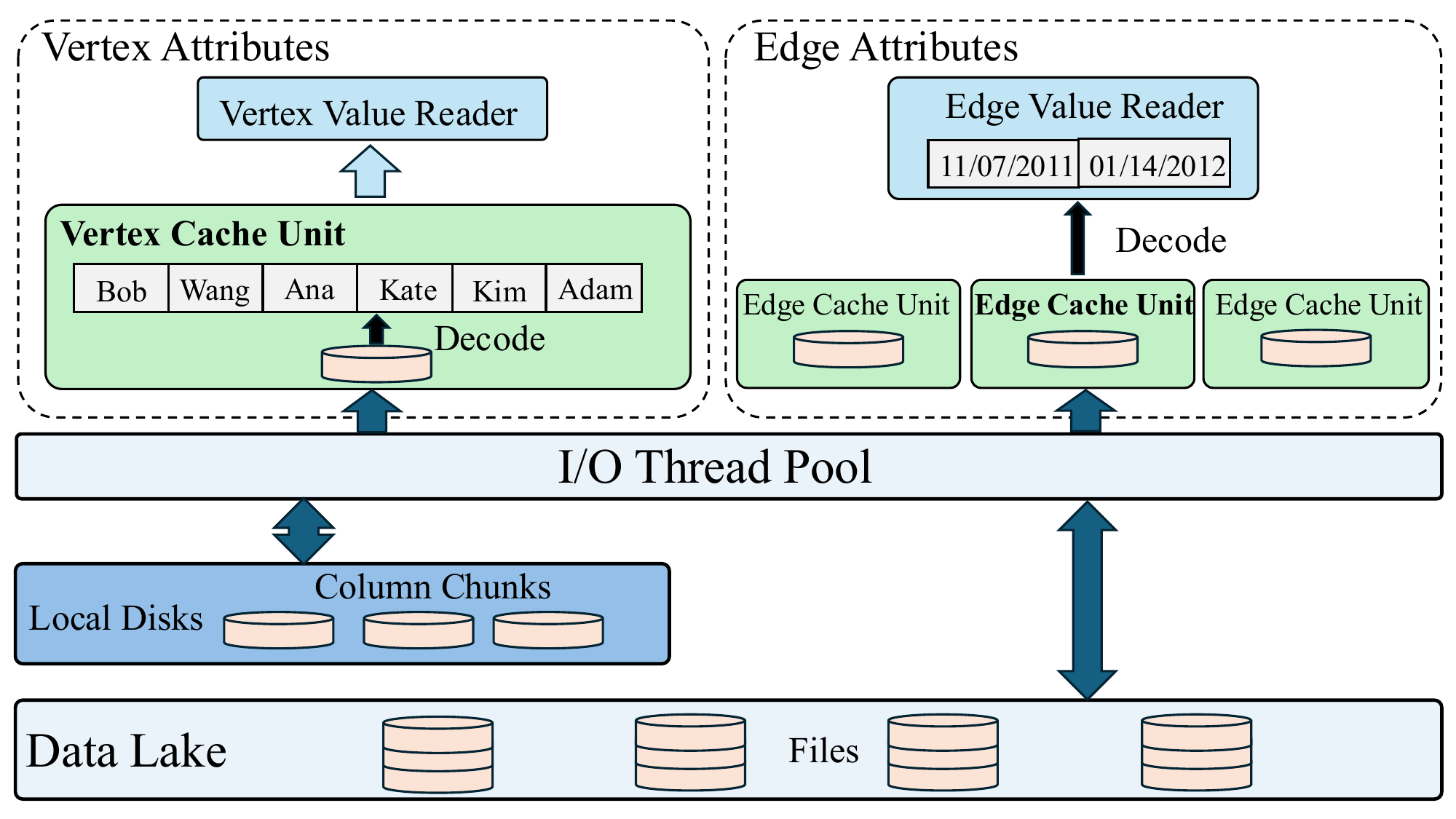}
\caption{Cache Management}\label{fig:CacheManagement}
\end{figure}

\subsection{Graph-Aware Cache Unit}\label{sec:cacheunit}


Next, we introduce how to optimize attribute access to accommodate graph access patterns.


Lakehouse column chunks are optimized for scanning. However, graph analytics relies on neighbor traversal using graph topology, thereby exposing an irregular access pattern for vertex attributes. In standard Lakehouse engines, this incurs frequent re-decoding of column chunks.
In addition, graph analytics often involves complex filter predicates that depend on properties from both edges and neighboring vertices (aka cross-entity predicates). This dependency necessitates row-level evaluation. Although edge attributes in \sys{} adopt a scan-oriented access pattern, evaluating these values individually still imposes significant decoding overhead. 

To address these issues, we introduce graph‑aware cache units. We cache decoded values from column chunks and implement specialized value readers to retrieve attribute values by row index during query processing. As illustrated in Figure~\ref{fig:CacheManagement}, these units can be classified into vertex cache units (e.g., \textit{Name} column chunks of \textit{Person} vertex table) and edge cache units (e.g., \textit{createDate} column chunks of \textit{Knows} edge table), each employing distinct caching strategies detailed below.


\myline{Vertex Cache Unit} 
During graph traversal, vertices are accessed irregularly, resulting in random access requests on vertex column chunks. To avoid re-decoding column chunks, we pre-allocate a decoded value array for each vertex column chunk. This array provides sufficient capacity to cache all values within the chunk, avoiding the performance overhead of dynamic resizing and data copying.

The decoded value array starts empty. During execution, an internal decoder populates it sequentially from the column chunk, ensuring the array maintains a contiguous prefix of decoded entries. For example, if the value array currently caches the first 100 values and the value reader requests the entry at index 300, the vertex cache unit decodes the values from index 100 through 300 and populates the intermediate entries, thereby preserving the contiguity of the prefix. 

We decode column chunks and ensure contiguity in the value array for two primary reasons: First, it prevents redundant decoding. Second, it simplifies status management by maintaining an incremental index boundary.

This array enables efficient point lookups for vertex attributes. For example, if a value reader requests the value at index 50, it can be retrieved directly via the array index. This functionality is important for irregular vertex access patterns.

For variable-length data types such as strings, we allocate separate memory blocks for the content and store only the corresponding offsets and lengths within the value array.


\myline{Edge Cache Unit} Cross-entity predicates in graph analytical queries necessitate row-level evaluation of edge attributes. To reduce amortized decoding overhead, we decode values from column chunks in batches to improve CPU cache locality. 

Value readers maintain a temporary buffer to cache values during edge attribute scanning.
For example, if a reader requires the value at index 50, it decodes the values from index 40 to 60, caches them in its temporary buffer, and retrieves the value at index 50. On subsequent requests, if the reader requires the value at index 55, it directly retrieves the value from the buffer, reducing the amortized decoding overhead. If a requested index falls outside the buffered range (e.g., index 100), the reader advances to the relevant page, decodes it in batches, and updates the buffer to include the required entry.

It is not worth adopting the decoded value array design in edge cache units. Our experiment in Section~\ref{comparewithcacheunit} demonstrates that for most OLAP queries, the performance gains of this approach are limited, while it incurs significant memory overhead due to the large volume of edges. Thus, a sliding-window design during edge scans best improves the OLAP workload efficiency.



\subsection{Priority-Based Cache Replacement}\label{sec:cacheallocation}



As illustrated in Figure~\ref{fig:CacheManagement}, we employ a two-tier cache architecture. We maintain a memory layer as a high-performance subset of a larger disk-based cache layer to balance access latency with storage capacity.
If edge cache units are evicted from memory, they are discarded because the underlying column chunks persist on local disks. In contrast, if vertex cache units are evicted from memory, their decoded values are flushed to disk to avoid redundant decoding during subsequent retrieval. Units evicted from local storage are directly deleted. We do not persist decoded vertex attributes back to the data lakes because it would incur unnecessary network overhead and create storage redundancy on the data lakes.

Next, we detail the cache eviction policy for memory and disk capacity constraints
SQL engines often treat column chunks uniformly. However, they do not distinguish between the vertex and edge cache units in \sys{}. First, vertex attributes are frequently accessed because vertices are the main access anchors, while edge attributes are accessed only once per traversal in \sys{}. Second, the overhead of evicting an edge cache unit is smaller because it can be discarded directly, whereas evicting a vertex cache unit incurs higher overhead because we have to flush decoded values to the disks.

To implement this differentiated eviction strategy, we employ a priority-aware sweep-clock algorithm (inspired by PostgreSQL), which provides a low-overhead approximation of LRU. Cache units are organized in a circular buffer through which a "clock hand" iterates. When a unit is accessed, its usage count is reset to a specific priority value: for instance, vertex units are assigned 3, while edge units are assigned 1. During each sweep, the clock hand decrements the usage count of the referenced unit. If the count reaches zero and the unit is not pinned by an active process, it is evicted to accommodate new data. This weighting mechanism naturally favors the retention of vertex units, which are costlier to re-decode and more likely to be reused.


\subsection{Prefetching}\label{sec:cacheprefetch}

Next, we describe our prefetching strategy for cache units during graph traversal.
We leverage an asynchronous I/O thread pool to fetch cache units from local disks or remote data lakes. However, identifying the optimal units to prefetch is still challenging.

Traditional SQL engines typically rely on sequential prefetching, loading subsequent column chunks into memory in advance. However, this approach is ill-suited for vertex cache units, which exhibit irregular access patterns that defy simple 'next-chunk' heuristics. While edge cache units are scanned sequentially which fits the ‘next chunk’ heuristic, prefetching all edge cache units still incurs a substantial I/O cost.

Thus, we propose a prefetching mechanism that relies on vertex frontiers during graph traversal and edge list statistics to anticipate data requirements.


\myline{Prefetching Based on Vertex Frontier} \sys{} evaluates graph queries by iteratively expanding a vertex frontier, a vertex set whose edges will be traversed (introduced in Section~\ref{sec:processing}). We use the vertex frontier as the first runtime signal for prefetching vertex cache units.

For each vertex file, we identify the Min-Max ID range encompassed by the current vertex frontier. We then iterate through row groups in this vertex file and compare each row group's ID range against the frontier's Min-Max values. When an overlap is detected, we prefetch the cache units for all columns required by the query. 

\myline{Prefetching Based on Edge List Statistics} Unselective prefetching of edge cache units incurs a substantial I/O cost. However, it is difficult to prune these edge cache units because each edge may have endpoints falling in the vertex frontier. 

We leverage statistical information from edge lists to prune edge cache units. Edge lists are logically split into multiple portions by row group index ranges of the edge files. For each edge list portion, we compute its Min and Max source IDs during edge list building. We then compare these ID ranges with the Min-Max ID ranges of the vertex frontier. Edge list portions and corresponding column chunks are pruned if no overlap is detected. This Min‑Max pruning is most effective when the underlying edge tables are sorted by the foreign key referring the source vertex.



\section{Lakehouse-Optimized Graph Analytics}\label{sec:processing}




In this section, we describe how \sys{} executes graph analytics over Lakehouse tables by leveraging the graph topology (in Section~\ref{sec:loading}) and cached graph properties (in Section~\ref{sec:cache}).


Consistent with the compute framework of TigerGraph~\cite{tigergraph}, \sys{} employs an \textit{accumulator-based aggregation paradigm}, executing graph analytics through declarative vertex state transitions within a BSP model. In this model, execution is strictly driven by an active vertex set. In each superstep, the engine executes either a \textit{VertexMap} or an \textit{EdgeMap} primitive. \textit{VertexMap} applies a UDF to the current active set with access to the static vertex properties and runtime accumulators; \textit{EdgeMap} applies a UDF to the edges incident to the active vertices. During edge processing, the engine reads properties and accumulators of an edge and its neighboring vertices (or endpoints), then updates the accumulators of both endpoints. The model enforces strict synchronization between steps, where the active set for the subsequent iteration is dynamically resolved: it may consist of a filtered subset of vertices following a \textit{VertexMap}, or the source (or target) endpoints of a filtered edge set processed during an \textit{EdgeMap}.

However, the execution of these primitives in TigerGraph relies on a \textit{proprietary storage layer}. It stores graph topology and graph content in a proprietary CSR structure. In contrast, Lakehouse tables utilize open formats, which decouples the storage layer from the query engine's direct management..

To address this challenge, \sys{} adapts the \textit{VertexMap} and \textit{EdgeMap} primitives for Lakehouse architectures. We rename \textit{EdgeMap} as \textit{EdgeScan} to emphasize its scan-based execution. These two Lakehouse-optimized primitives seamlessly integrate into GSQL queries by providing the same semantics as the original \textit{VertexMap} and \textit{EdgeMap}.

Next, we explain how to efficiently execute these two primitives in Section~\ref{sec:processingoperator}. We then explain their execution in a distributed environment in Section~\ref{sec:processingdistributed}.

We use a representative graph OLAP query from the LDBC\_SNB BI~\cite{ldbc-snb}, a well-known graph analytics benchmark, as an example to introduce \sys{} primitives. This query finds women who have created comments with the tag "Music" after 2010-01-01 and aggregates the number of comments each such person creates. This GSQL query (shown below) specifies the pattern in the FROM clause, where vertex types ( \textit{Person} and \textit{Comment}) are in the parenthesis, and edge types ( \textit{HasTag} and \textit{HasCreator}) are in the square brackets. The number of comments each person creates is stored in accumulator \textit{@sum}.

\begin{lstlisting}[basicstyle=\fontsize{5.7pt}{6.7pt}\ttfamily]
SELECT p
FROM (t:Tag) <- [e1:HasTag] - (c:Comment) 
- [e2:HasCreator] -> (p:Person)
WHERE t.name == "Music" 
    AND e2.date > Date('2010-01-01')
    AND p.gender == "Female"
ACCUM p.@sum += 1;
\end{lstlisting}

\subsection{Lakehouse-Optimized Parallel Primitives}\label{sec:processingoperator}



\myline{VertexMap} 
This primitive takes an active vertex set as well as a UDF for vertices as input. The active vertex set is a highly compressed bitmap, segmented based on vertex files. The output of this primitive is a vertex subset that passes filtering. For example, one \textit{VertexMap} for the example query filters all \textit{Tag} vertices based on the \textit{name} attribute, and output vertices that satisfy the predicate \textit{t.name == "Music"}.

\textit{VertexMap} concurrently evaluates vertices across multiple vertex files. It creates value readers for all involved vertex columns and assigns thread tasks for vertex files that have active vertices. For each vertex file, it iterates over the bitmap of the vertex file and feeds value readers with transformed vertex IDs. As mentioned in Section~\ref{sec:cacheunit}, these value readers can retrieve attribute values from cache units given the file path and row index, which can be decoded from our transformed vertex IDs. A vertex row is fully materialized as value readers successfully retrieve the vertex properties from cache units. Then \textit{VertexMap} applies the UDF on this vertex.

\myline{EdgeScan} 
This primitive takes a source vertex set as well as UDFs for source vertices, neighboring edges, and target vertices. After execution, it outputs the source or the target vertex set of filtered edges. 
\textit{EdgeScan} utilizes an \textbf{edge-centric processing method} by leveraging edge lists and Lakehouse cache units, rather than the vertex-centric approach used by TigerGraph.


Specifically, \textit{EdgeScan} concurrently scans edge lists and applies UDFs for each edge incident to the active vertex set and its endpoints. 
It first creates value readers upon vertex/edge properties that are involved (projected or filtered) in the query, such as \textit{date} of \textit{HasCreator} in the example. Then \textit{EdgeScan} uses multiple threads to scan edge lists and creates thread tasks for each edge list. 

A thread is assigned to iteratively scan an edge list. For each entry, the system verifies if the source vertex ID exists within the input set. Upon a match, it decodes the file paths and row indices for both endpoints from the edge entry, while simultaneously calculating the edge’s position based on its offset. These coordinates are then passed to the value readers to retrieve the property values for both the vertices and the edge itself. 
Thus, for these edges, we fully materialize the neighboring vertex rows and edge rows since all value readers retrieve their properties from cache units, as shown in Figure~\ref{fig:triplet}. Then we apply UDFs on these edge rows and vertex rows, evaluating filter predicates and updating accumulators. For instance, the example query has two \textit{EdgeScan} primitives for two-hop graph traversal. The second \textit{EdgeScan} evaluates the filter predicate \textit{e2.date > date ('2010-01-01')} from the \textit{WHERE} clause and updates the accumulator \textit{p.@sum} from the ACCUM clause for each edge and its endpoints. The engine sends these accumulator updates to the endpoints and these updates are combined before the next iteration.

We also apply pruning when scanning edge lists whenever possible. For example, we prune irrelevant edge lists based on specified edge types, which determine the edge tables and underlying edge files that need to be considered. In addition, we prune edge lists by comparing their min-max vertex ID ranges with the vertex frontier, as described in Section~\ref{sec:cacheprefetch}.

\begin{figure}[tbp]
\centering
\includegraphics[width=0.47\textwidth]{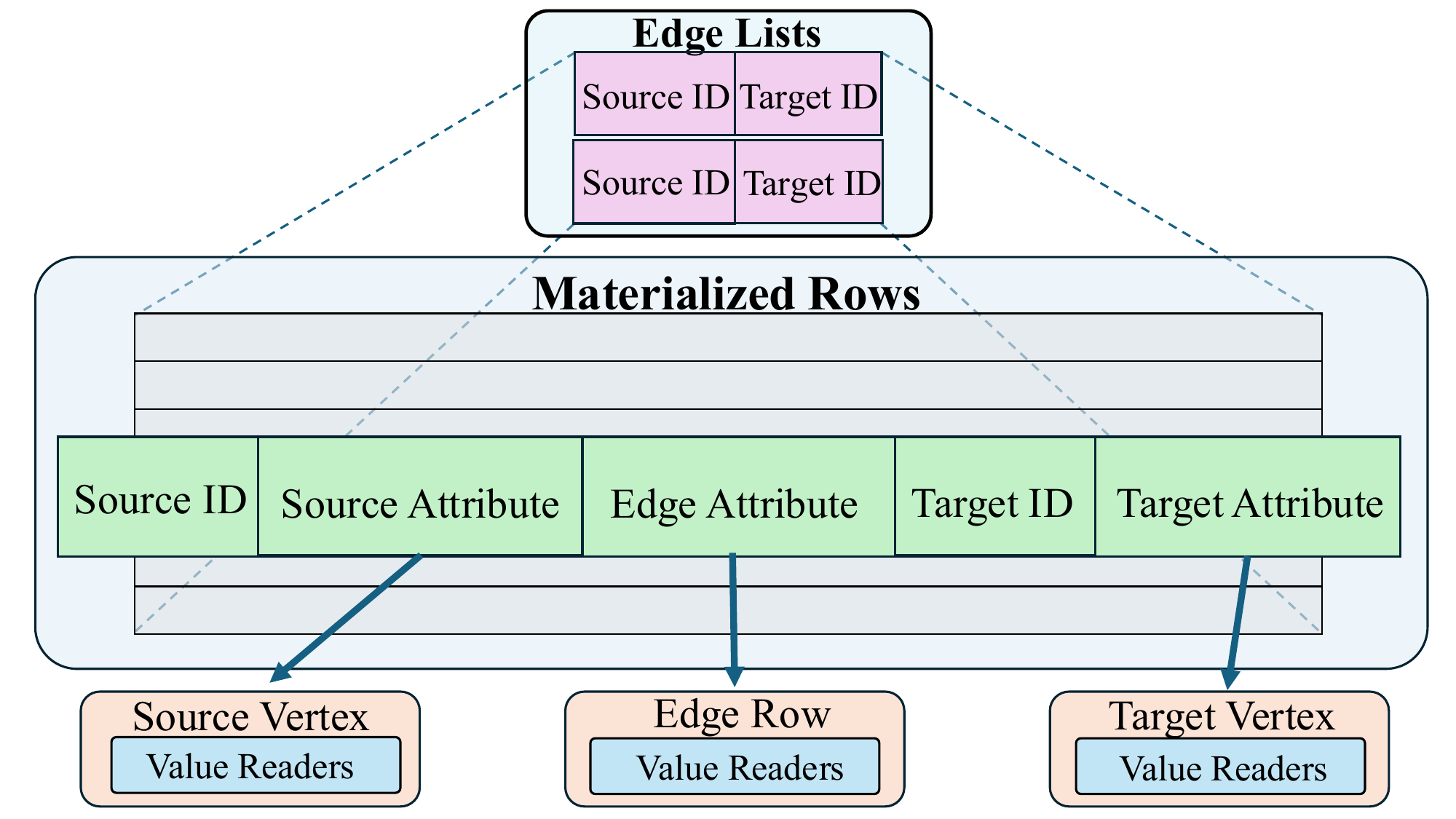}
\caption{Vertex and Edge Row Materialization}\label{fig:triplet}
\end{figure}

For queries that require traversing incoming edges rather than outgoing edges, \textit{EdgeScan} supports bidirectional traversal without extra storage cost. When traversing outgoing edges, we treat the first vertex ID in an edge list entry as the source vertex and the second as the target. When traversing incoming edges, we treat the first ID as the target vertex and the second as the source. In contrast, CSR-based engines such as TigerGraph usually need to store edges twice to support bidirectional traversal.




\subsection{Distributed Query Processing}\label{sec:processingdistributed}

Next, we introduce how to efficiently execute the two new primitives to evaluate distributed graph queries. We mainly focus on \textit{EdgeScan} because the distributed execution of \textit{VertexMap} is straightforward: every compute node processes their owning vertices.

As mentioned in Section~\ref{sec:optimizedgraphtopo}, we take a file-based sharding method that randomly partitions vertex and edge files to different compute nodes. However, this method incurs a problem for distributed execution of \textit{EdgeScan}: An edge may not co-locate with its endpoints.

There are two naive solutions. The first solution is to directly access remote data. Each time \textit{EdgeScan} evaluates an edge and encounters remote endpoints, it directly requests vertex properties from remote compute nodes. However, this solution will trigger massive requests and bottleneck \textit{EdgeScan} with network latency. The second solution is to replicate the vertices by allowing all compute nodes to access the vertex files. However, this solution increases memory burden and leads to redundant decoding over vertex cache units. 

Notice that we replicate the Vertex IDM in Section~\ref{sec:optimizedgraphtopo} because it is relatively small and significantly improves the edge list building. While we do not replicate vertex attributes here because the network latency is tolerable after batching requests.



Thus, we propose a two-pass scanning approach, where the compute nodes still request remote vertex data to reduce memory and decoding burden, but batch these requests to minimize overall latency. 
In the first pass, each compute node scans local edge lists to identify which vertex rows should be materialized but are located on remote nodes, such as P1 and P6 in the example of Figure~\ref{fig:edgeprocess}. Then it collects all remote data requests and sends them in one batch. Remote nodes filter the vertex data before returning them in the fashion of the classic filter push down technique (e.g.,  P7 in Figure~\ref{fig:edgeprocess} is filtered out by the predicate \textit{p.gender == "Female"}). The active vertex set is refreshed, only recording filtered vertices from remote nodes. In the second pass, each compute node evaluates on materialized vertex and edge rows as mentioned in Section~\ref{sec:processingoperator}. The partial updates of vertex accumulators are pushed back to the host machines at the end to be combined into final values.

\begin{figure}[tbp]
\centering
\includegraphics[width=0.47\textwidth]{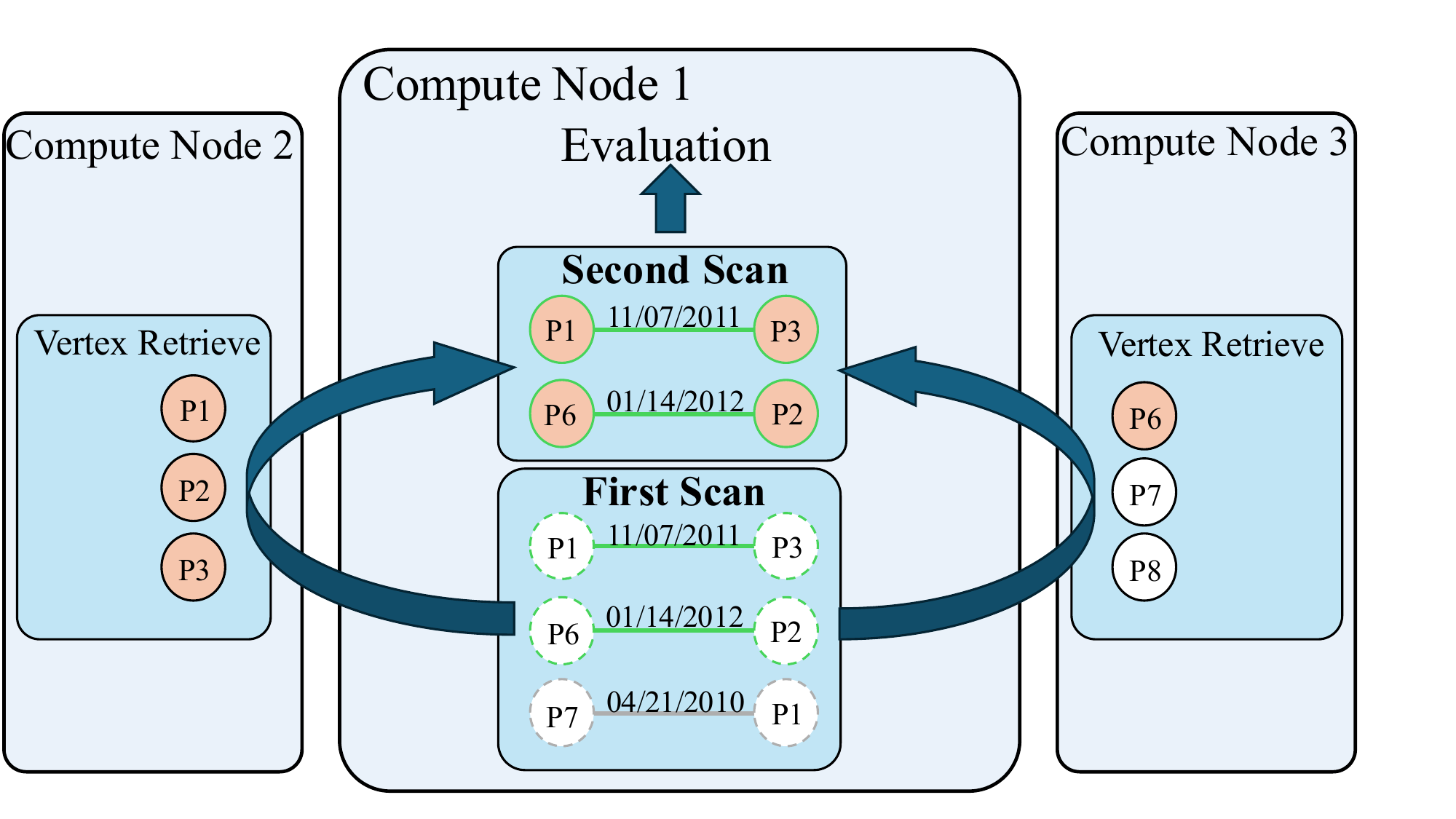}
\caption{Distributed \textit{EdgeScan}}\label{fig:edgeprocess}
\end{figure}

\section{Experiments}\label{sec:experiment}


In this section, we present experimental results to evaluate \sys{}. 

\subsection{Experiment Setting}\label{sec:expsetting}

\myline{Experimental Platform} 
We use Apache Iceberg~\cite{Iceberg}, one of the most popular Lakehouse systems, in our experiments. We store the Iceberg tables in the AWS S3 us-east-2 region. By default, we perform experiments on a Linux server with a 40-core Intel Xeon CPU (2.30 GHz), 64 GB DRAM, and a 1.8 TB NVMe SSD, with 1.1 GB/s network throughput to AWS S3. Distributed experiments are performed using AWS EC2 instances in the same region (Intel Xeon Platinum 8175M, 32 vCPUs, 128 GB memory, and 64 MB L3 cache).



\myline{Datasets} 
We mainly perform our experiments with LDBC\_SNB BI~\cite{ldbc-snb}, a standard workload for graph analytics. It is a social network graph consisting of vertices and edges such as \textit{Comment}, \textit{Person}, and \textit{Comment\_HasCreator\_Person}. This workload has datasets of different scales, whose sizes scale proportionally with the scale factor (SF), as shown in Table~\ref{tab:dataset}. For example, SF1 has 3~M vertices and 17~M edges, while SF30 has a total of 78~M vertices and 505~M edges. In our experiments, we use SF1 to SF300 for single-machine tests and SF30 to SF1000 for scalability tests. We pre-process these datasets by loading them into Lakehouse tables, where each vertex or edge type corresponds to a table. To best evaluate the concurrency of \sys{}, each table is split into 32 data files, since 32 vCPUs is a common configuration for many compute nodes. When SF is large, the file size of some tables may be too large. In this case, we split the table into more files to ensure that each file is between 128~MB and 256~MB, a recommended data size range in the AWS guide~\cite{aws_guide}.

Since graph algorithms are also an important part of graph analytics, we conduct graph algorithm experiments with LDBC GraphAnalytics~\cite{graphanalytics}, another popular graph algorithm benchmark from LDBC. Specifically, we use Graph500-22, a representative dataset from this benchmark containing 2.4~M vertices and 64.2~M edges. All five graph algorithms in Graph500-22 are evaluated, including PageRank (PR), weakly connected components (WCC), community detection using label propagation (CDLP), local clustering coefficient (LCC), and breadth-first search (BFS).

\begin{table}[tbp]
\small
\centering
\caption{\textbf{Statistics of LDBC\_SNB}}\label{tab:dataset}
\resizebox{0.8\linewidth}{!}{%
\begin{tabular}{r|r|r|r}\hline\hline
\textbf{Scale Factor} & \textbf{\# Vertex Count} & \textbf{\# Edge Count} & \textbf{Size (GB)}\\\hline\hline
SF1 & 2,997,352 & 17,196,776 & 0.28\\\hline
SF3 & 8,513,157 & 51,035,227 & 0.80\\\hline
SF10 & 27,231,349 & 170,343,945  & 2.62 \\\hline
SF30 & 78,244,709 & 505,722,361  & 7.79 \\\hline
SF100 & 254,634,696 & 1,703,042,944  & 26.76 \\\hline
SF300 & 738,162,716 & 5,078,844,191  & 79.94 \\\hline
SF1000 & 2,433,117,531 & 17,203,259,133  & 278.46 \\\hline\hline
\end{tabular}%
}
\end{table}

\myline{Competitors} 
We mainly compare \sys{} with PuppyGraph~\cite{PuppyGraph}, which is the state-of-the-art graph compute engine that supports the Lakehouse. Since \sys{} is built on TigerGraph, we also compare it with TigerGraph to evaluate the performance loss incurred when supporting the Lakehouse (Section~\ref{comparewithtigergraph}).




\subsection{Results on Startup Loading Time}\label{sec:exploading}

In this experiment, we evaluate the startup time of \sys{} in comparison with PuppyGraph. We distinguish between the startup time for the first connection, where the graph topology has not been built, and the startup time for the second and subsequent connections (referred to as the "second connection" for brevity), where the graph topology has been persisted in the data lake.

For PuppyGraph, since it is not open source and we cannot directly check the engine status as we can for \sys{}, we measure its startup latency by periodically sending a lightweight graph query until it successfully returns results. This query is a simple one-hop query over the \textit{Tag} and \textit{TagClass} vertex types, both of which have a small number of vertices.


Figure~\ref{fig:startup_time} shows the results of the startup time evaluation. These results demonstrate that \sys{} has much lower startup time than PuppyGraph. It achieves \textbf{1.7}$\times$ to \textbf{4.0}$\times$ faster startup for the first connection because \sys{} only loads the graph topology. After persisting the built topology in data lakes, it achieves \textbf{6.9}$\times$ to \textbf{26.3}$\times$ faster startup for second and subsequent connections. These results show that \sys{} achieves significantly faster startup than PuppyGraph.

Figure~\ref{fig:startup_pies} presents the breakdown of \sys{}'s startup time on LDBC\_SNB SF100. The breakdown of the first connection shows that the startup time is dominated by graph topology building, where Vertex IDM building takes 21.9\% and edge list building takes 54.5\%. The breakdown of the second connection (and subsequent connections) shows that the startup time is dominated by setting up connections with the Lakehouse and data lakes, which incurs nearly constant time. This is because we can skip graph topology building after persisting edge lists in the data lakes.


\begin{figure}[tbp]
\centering
\includegraphics[width=0.47\textwidth]{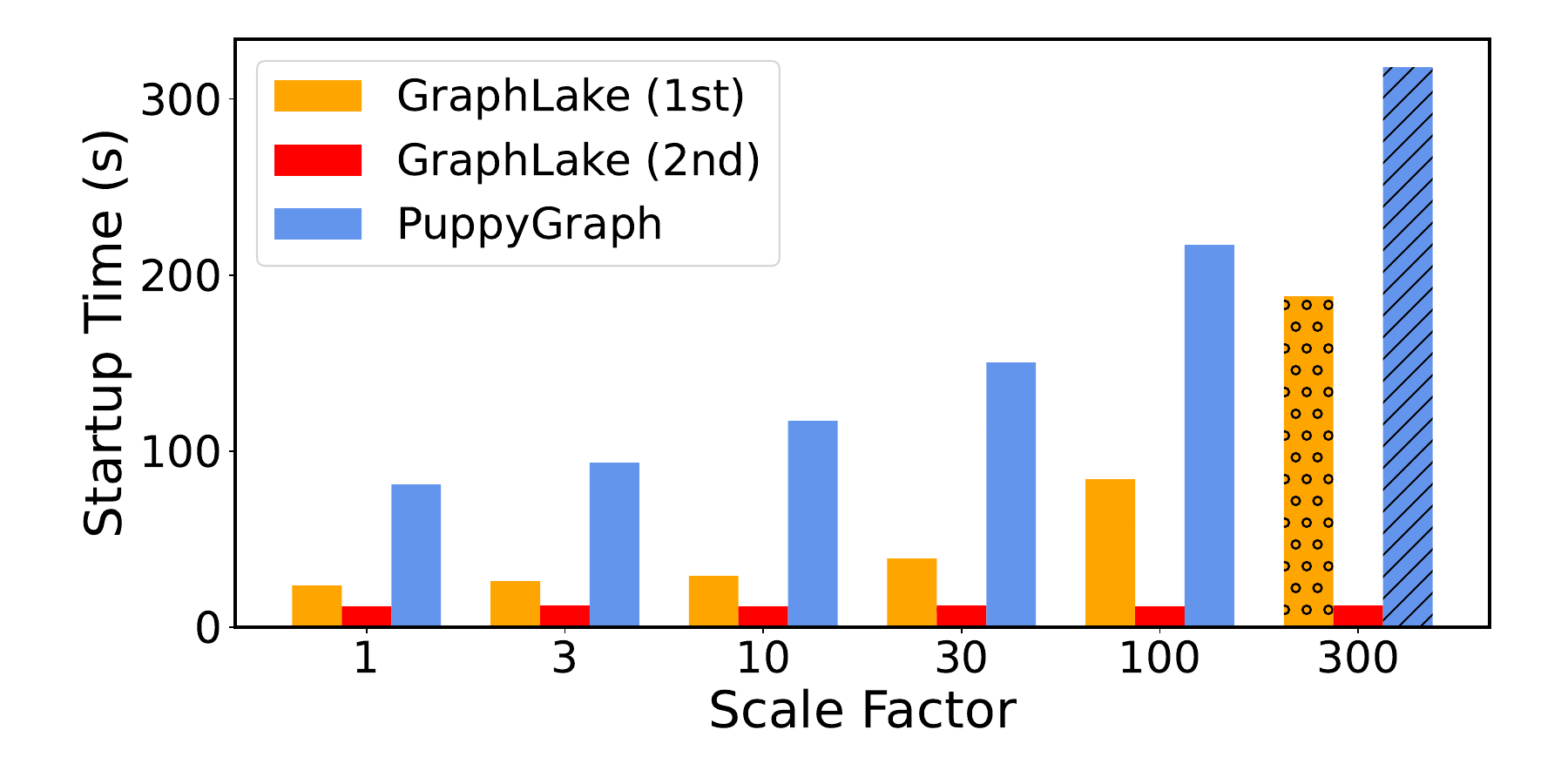}
\caption{Startup Time Evaluation}
\label{fig:startup_time}
\end{figure}

\begin{figure}[tbp]
\centering
\includegraphics[width=0.47\textwidth]{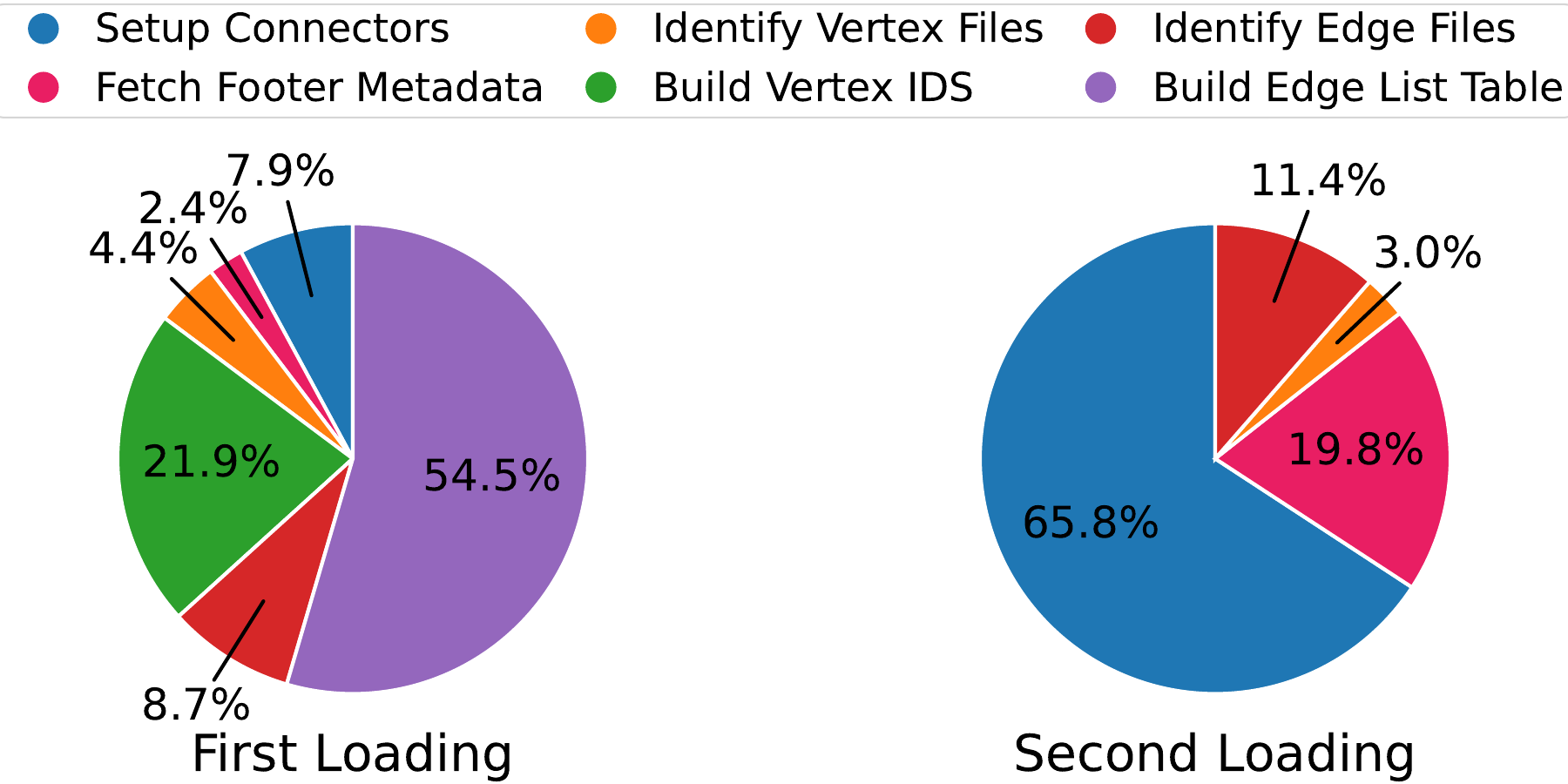}
\caption{Startup Time Breakdown on SF100}
\label{fig:startup_pies}
\end{figure}

\subsection{Results on Graph-Aggregation Queries}\label{sec:expquery}

In this experiment, we evaluate the graph-aggregation query performance of \sys{} in comparison with PuppyGraph. We pick five representative queries from the standard LDBC benchmark~\cite{LDBCGITHUB}, which provides Cypher queries for Neo4j and GSQL queries for TigerGraph. We adopt these queries but slightly modify the Cypher queries because some syntax is not supported in PuppyGraph. We use the same query parameters to ensure a fair comparison.

We measure the query time for hot runs, when all data have been warmed up in memory. We also measure the query time for S3 and disk cold runs, which means that cache units and edge lists are loaded from S3 or local disks when queries start. Since PuppyGraph has a background process that automatically fetches data from the Lakehouse to local disks, we cannot present its S3 cold run results.

Figure~\ref{fig:sf30_latency} and Figure~\ref{fig:sf300_latency} present query times for both hot runs and cold runs. For experiments on SF30, we observe that \sys{} is up to \textbf{60.3}$\times$ faster than PuppyGraph for hot runs and up to \textbf{29.8}$\times$ faster for disk cold runs. For experiments on SF300, \sys{} is up to \textbf{12.8}$\times$ faster than PuppyGraph for hot runs and up to \textbf{4.4}$\times$ faster for disk cold runs. PuppyGraph fails on BI8 due to out-of-memory (OOM), but \sys{} can still execute it within a reasonable time. This is because our engine directly leverages edge lists and Lakehouse-native cache units to efficiently traverse the graph.

The above results show that \sys{} consistently matches or outperforms PuppyGraph across different graph-aggregation queries. In several workloads, \sys{} achieves an order-of-magnitude lower query time for both hot and disk cold runs.

\begin{figure}[tbp]
\centering
\includegraphics[width=0.47\textwidth]{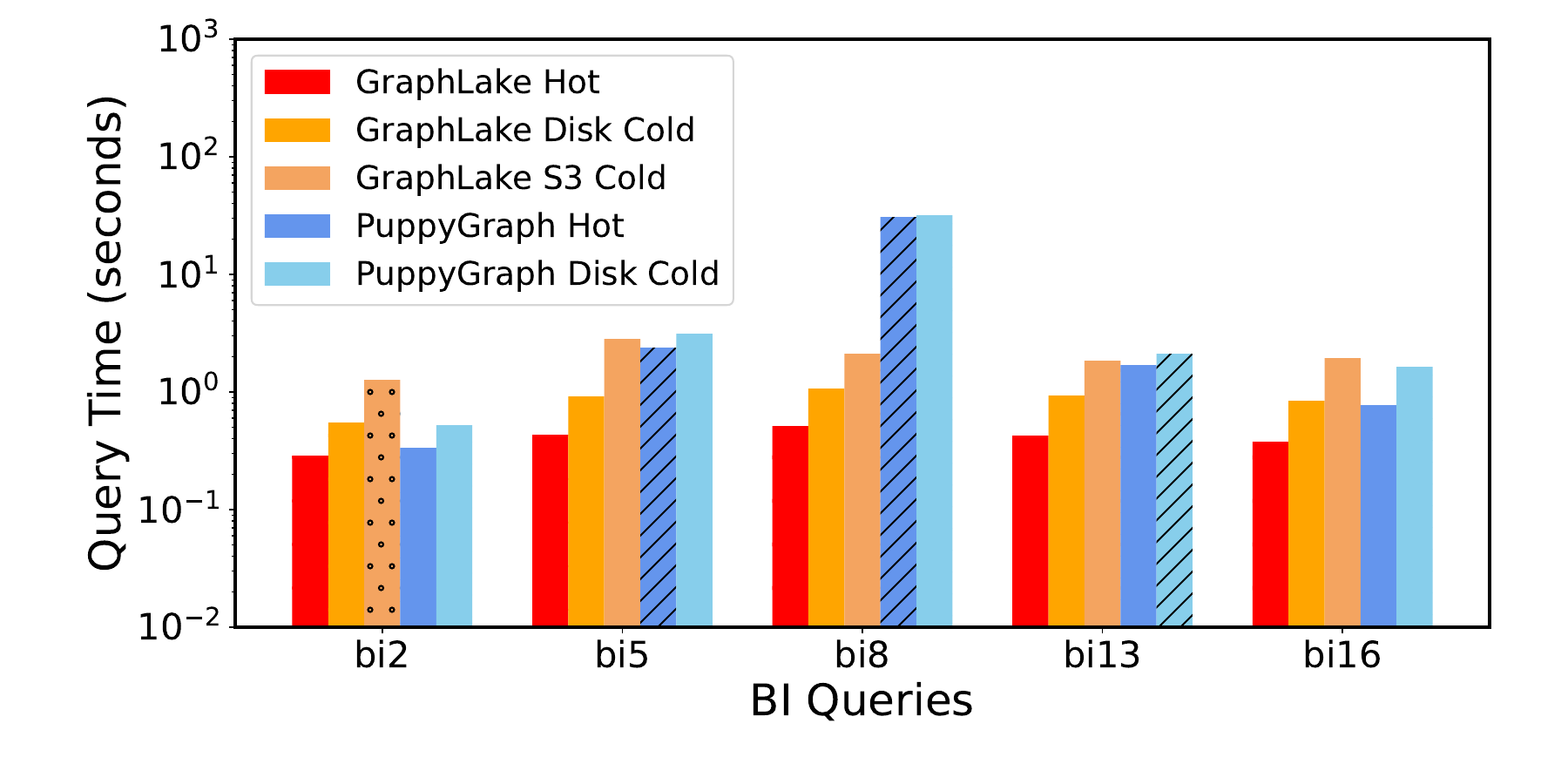}
\caption{Query Time Evaluation for SF30}
\label{fig:sf30_latency}
\end{figure}

\begin{figure}[tbp]
\centering
\includegraphics[width=0.47\textwidth]{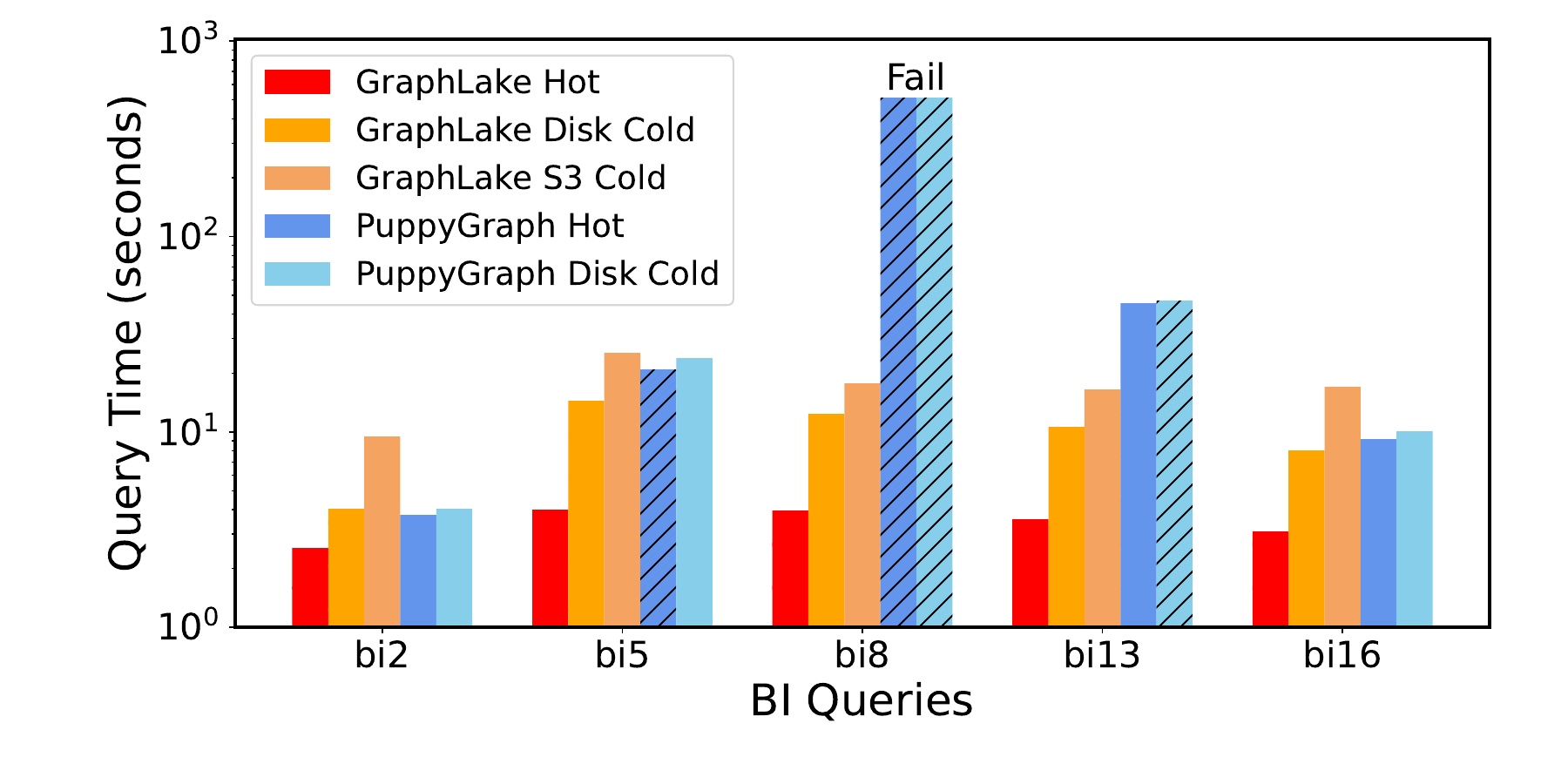}
\caption{Query Time Evaluation for SF300}
\label{fig:sf300_latency}
\end{figure}

\subsection{Results on Graph Algorithms}\label{sec:expalgo}

In this experiment, we compare the running time of graph algorithms in \sys{} with that of PuppyGraph. We leverage their built-in procedures to execute graph algorithms on the Graph500-22 dataset with identical algorithm parameters.

As shown in Table~\ref{tab:graph_algorithms}, for PR and WCC, \sys{} is up to \textbf{9.5}$\times$ faster than PuppyGraph. For CDLP, BFS, and LCC, \sys{} can execute these algorithms in a reasonable time, while PuppyGraph either fails due to OOM or does not provide a procedure to support the algorithm. This is because the vertex accumulators in GSQL efficiently reduce intermediate results, while PuppyGraph may encounter OOM failures due to overly large intermediate working tables.



\begin{table}[tbp]
\small
\centering
\caption{\textbf{Graph Algorithm Running Time}}\label{tab:graph_algorithms}
\resizebox{0.8\linewidth}{!}{%
\begin{tabular}{c|c|c|c|c|c}\hline\hline
\textbf{} & \textbf{PR} & \textbf{WCC} & \textbf{CDLP}& \textbf{LCC} & \textbf{BFS}\\\hline\hline
\sys{} & \textbf{28.0s} & \textbf{9.3s} & \textbf{117.0s} & \textbf{640.8s} & \textbf{3.5s}\\\hline
PuppyGraph & 59.6s & 88.1s & OOM & Unsupported & OOM\\\hline\hline
\end{tabular}%
}
\end{table}

\subsection{Scalability Evaluation}\label{sec:expscala}

In this experiment, we evaluate the scalability of \sys{} in terms of the number of nodes and the dataset size (i.e., scale factor). Since Lakehouse files are partitioned across nodes, we present throughput to evaluate query performance so that each node can fully leverage its compute resources.

We use an additional sender machine that evenly distributes requests across all machines to minimize the impact of network ports. The sender machine uses a popular HTTP benchmarking tool, wrk2~\cite{wrk2}. In this experiment, it maintains 16 connections and forks 2 threads in total because OLAP queries are relatively long-running. Each thread prepares payloads with randomized query parameters to minimize the impact of CPU caches caused by identical payloads. The number of requests sent per second is high enough to saturate the throughput of our system.

Figure~\ref{fig:single_throughput} presents how throughput scales with larger datasets on a single node. We present the throughput of \sys{} for two queries, BI2 and BI16. As the scale factor increases from SF1 to SF300, the throughput of BI2 decreases from 113.6 to 0.6 queries per second, corresponding to about a \textbf{186}$\times$ drop for a \textbf{300}$\times$ increase in data size. For BI16, throughput drops from 128.8 to 0.6 queries per second over the same range, roughly a \textbf{222}$\times$ decrease. Both results indicate good scalability. This is because scanning edge lists, the core evaluation step, achieves better CPU cache efficiency on larger datasets.

\begin{figure}[tbp]
\centering
\includegraphics[width=0.38\textwidth]{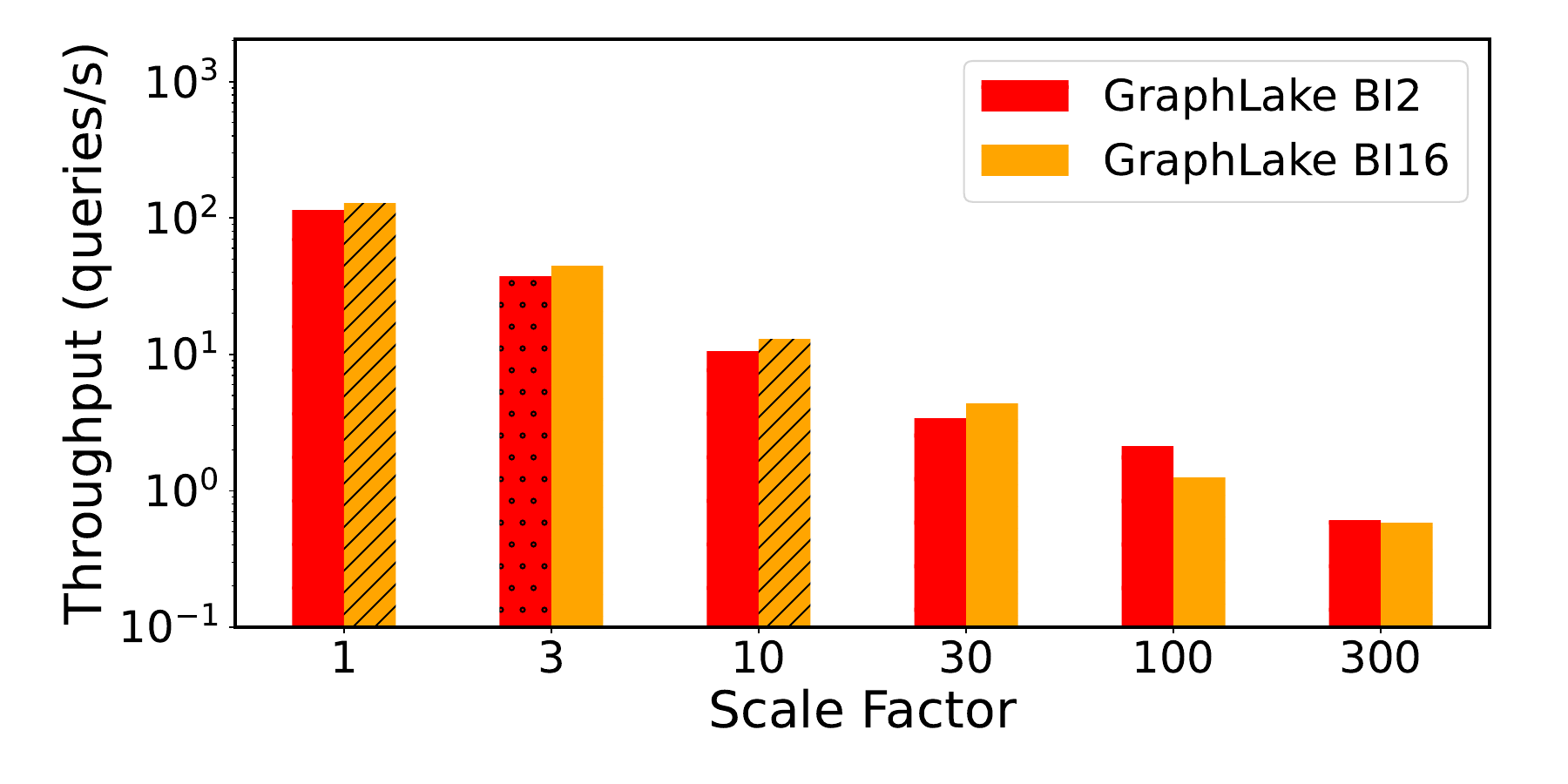}
\caption{Throughput Evaluation for BI2 and BI16}
\label{fig:single_throughput}
\end{figure}

Figure~\ref{fig:scala_startup_time} presents how the startup time scales with more machine nodes and larger datasets. Regarding the first startup, we can observe that \sys{} achieves close to linear scalability when the number of nodes is small. For example, when scaling from 1 node to 2 nodes, it achieves \textbf{1.8}$\times$ to \textbf{1.9}$\times$ faster startup across datasets of different sizes. This is because building edge lists, which can be efficiently scaled out with more nodes, still occupies the majority of the startup time. As the number of nodes increases and building edge lists takes relatively less time, \sys{} shows reduced scalability since Vertex IDM building is not partitioned. The scalability of the second startup is always poor since the majority of the time is spent setting up connections, which incurs approximately constant time.

Figure~\ref{fig:scala_bi_throughput} presents how the throughput of two queries, BI2 and BI16, scales with more machine nodes and larger datasets. We can observe that for larger datasets, \sys{} has better scalability. For example, on SF1000, \sys{} gains \textbf{1.8}$\times$ to \textbf{1.9}$\times$ higher throughput when doubling the number of nodes. However, on SF100, \sys{} only gains \textbf{1.2}$\times$ to \textbf{1.3}$\times$ higher throughput. This is because larger datasets lead to relatively more local computation. Although distributed query evaluation involves fetching vertex data from remote nodes, the amount of such data is still small compared to the size of local edge data. We can also observe that for a specific dataset size, scalability becomes worse as the number of nodes increases, since network communication becomes more complex and incurs relatively more overhead.

The above results show that \sys{} demonstrates strong scalability for both startup loading and query processing.

\begin{figure}[tbp]
\centering
\includegraphics[width=0.47\textwidth]{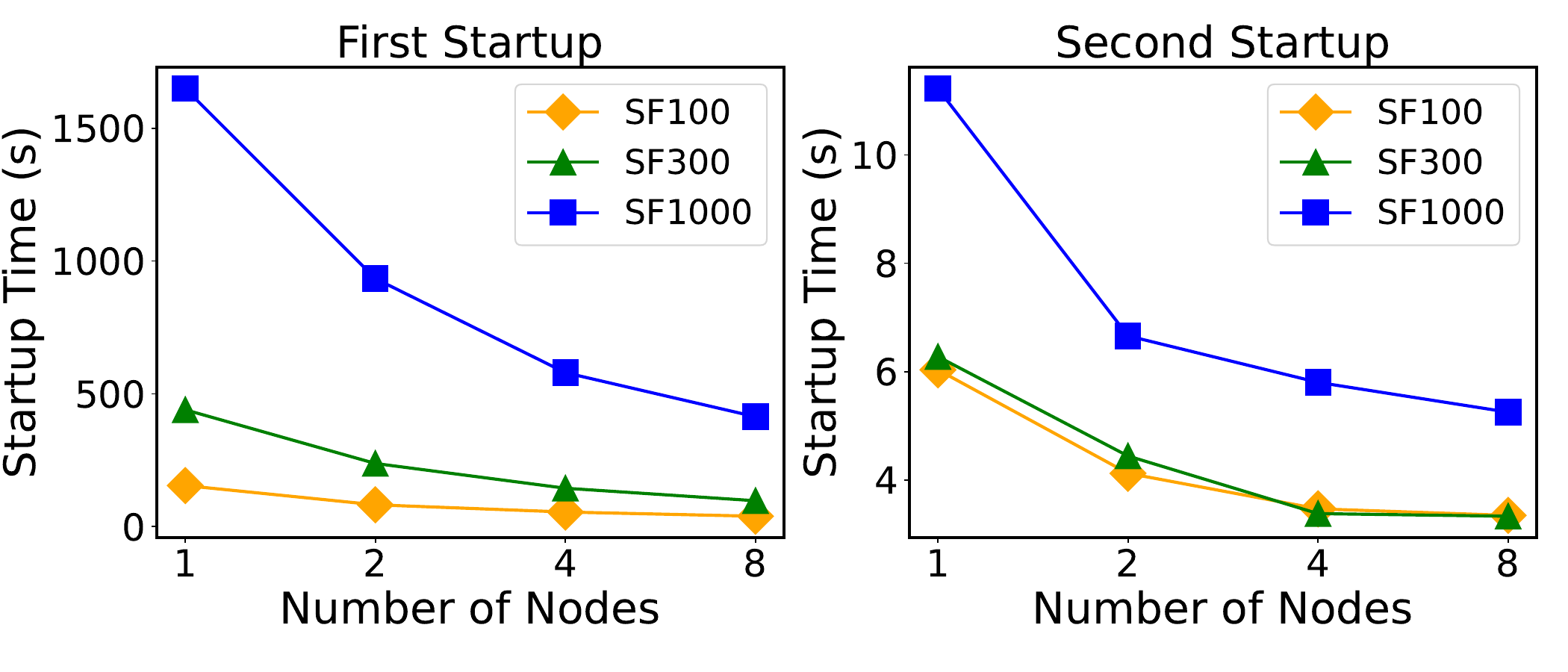}
\caption{Startup Time Scalability Evaluation}
\label{fig:scala_startup_time}
\end{figure}

\begin{figure}[tbp]
\centering
\includegraphics[width=0.47\textwidth]{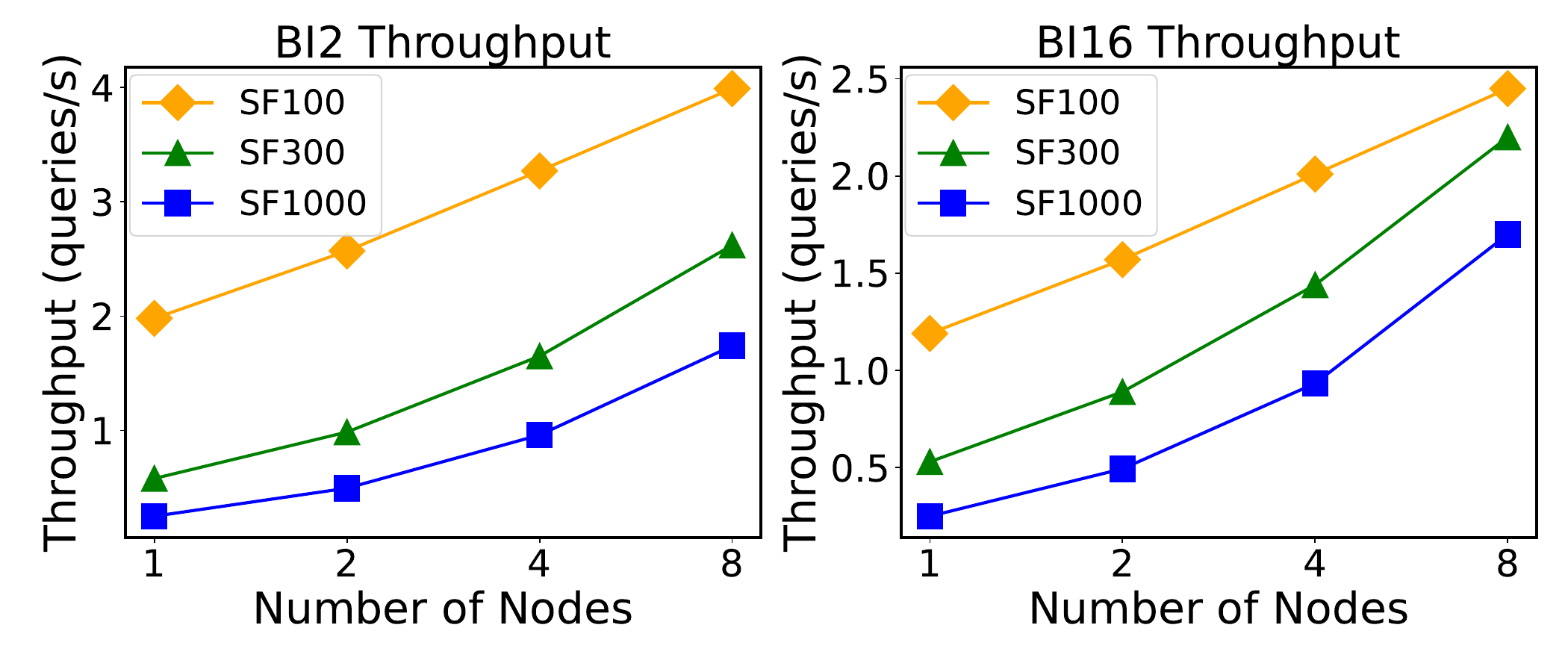}
\caption{Throughput Scalability Evaluation for BI2 and BI16}
\label{fig:scala_bi_throughput}
\end{figure}

\subsection{Additional Experiments}\label{sec:additional_exp}


Next, we conduct experiments to further analyze the performance of \sys{}.

\subsubsection{Comparing with TigerGraph}\label{comparewithtigergraph}


It is interesting to compare the query performance of \sys{} with TigerGraph because \sys{} is built on TigerGraph, although its performance is expected to be lower since \sys{} accesses attributes from open-format Lakehouse tables while TigerGraph has proprietary storage.

Because the main difference between these two systems lies between the implementation of the primitive \textit{EdgeMap} and \textit{EdgeScan}, we directly compare the execution of \textit{EdgeMap} and \textit{EdgeScan} under different selectivities. Specifically, we run a representative one-hop pattern match query on the LDBC\_SNB SF100 dataset; this query aims to find all persons who create given comments. This pattern match is common in LDBC\_SNB BI queries. We measure the execution time of \textit{EdgeMap} for the two systems. The selectivity is adjusted by the size of the input vertex set. To avoid the impact of row-oriented storage in TigerGraph and column-oriented cache units in \sys{}, this query does not involve attributes but applies a compute function on edges.

From the results in Figure~\ref{fig:in_memory_study}, we observe that \sys{} is even faster than TigerGraph when selectivity is greater than 10\%, since the sequential scanning of edge lists provides better CPU cache locality. However, when selectivity is small, for example, 0.01\%, TigerGraph can finish the execution in several milliseconds because it prunes many edges by vertex, while \sys{} spends hundreds of milliseconds scanning edge lists. Thus, we conclude that \sys{} best suits heavy OLAP workloads with high edge selectivity.

\begin{figure}[tbp]
\centering
\includegraphics[width=0.35\textwidth]{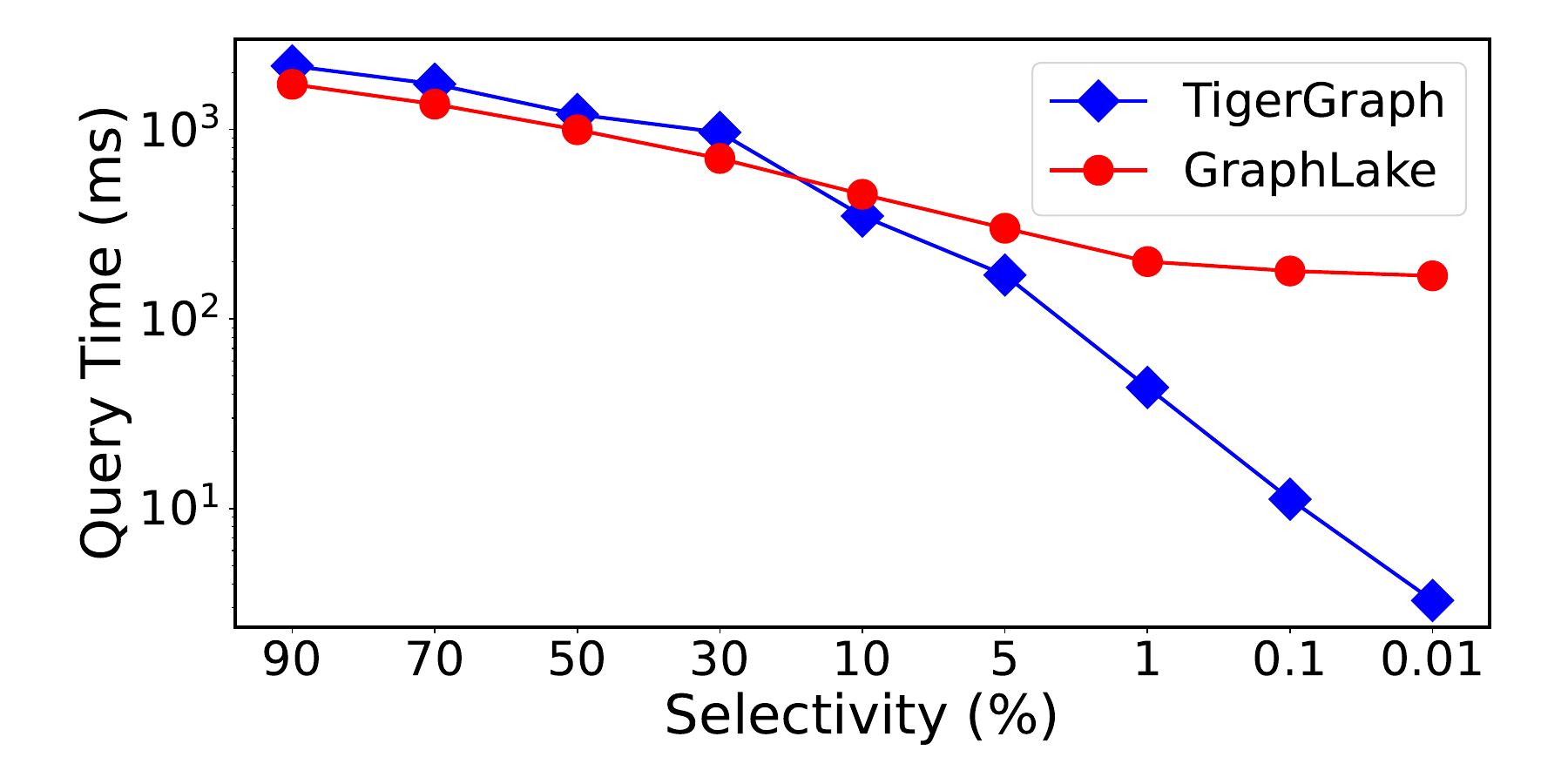}
\caption{Edge-centric processing using edge lists vs. vertex-centric processing using CSR structure}
\label{fig:in_memory_study}
\end{figure}

\subsubsection{Evaluating Graph-Aware Cache Unit}\label{comparewithcacheunit}

Graph-aware cache unit design decodes vertex column chunks and caches properties in a decoded value array, while a naive cache unit design repeatedly scans Lakehouse column chunks in a vectorized fashion. We want to compare the gap between these two designs.

We use the same query as in Section~\ref{comparewithtigergraph}, but add filters on vertex properties. We only measure the hot run query time to minimize I/O impact.

From results in Figure~\ref{fig:graph_aware_cache_unit_study}, we observe that the naive design exhibits very high query time, ranging from 1734.2 s to 2.5 s, because vertex column chunks are repeatedly decoded. In contrast, our graph-aware design has low query time, ranging from 357.1 ms to 148.2 ms across the same selectivity range. This study shows that our graph-aware cache units are highly effective for irregular graph access patterns.

\begin{figure}[tbp]
\centering
\includegraphics[width=0.35\textwidth]{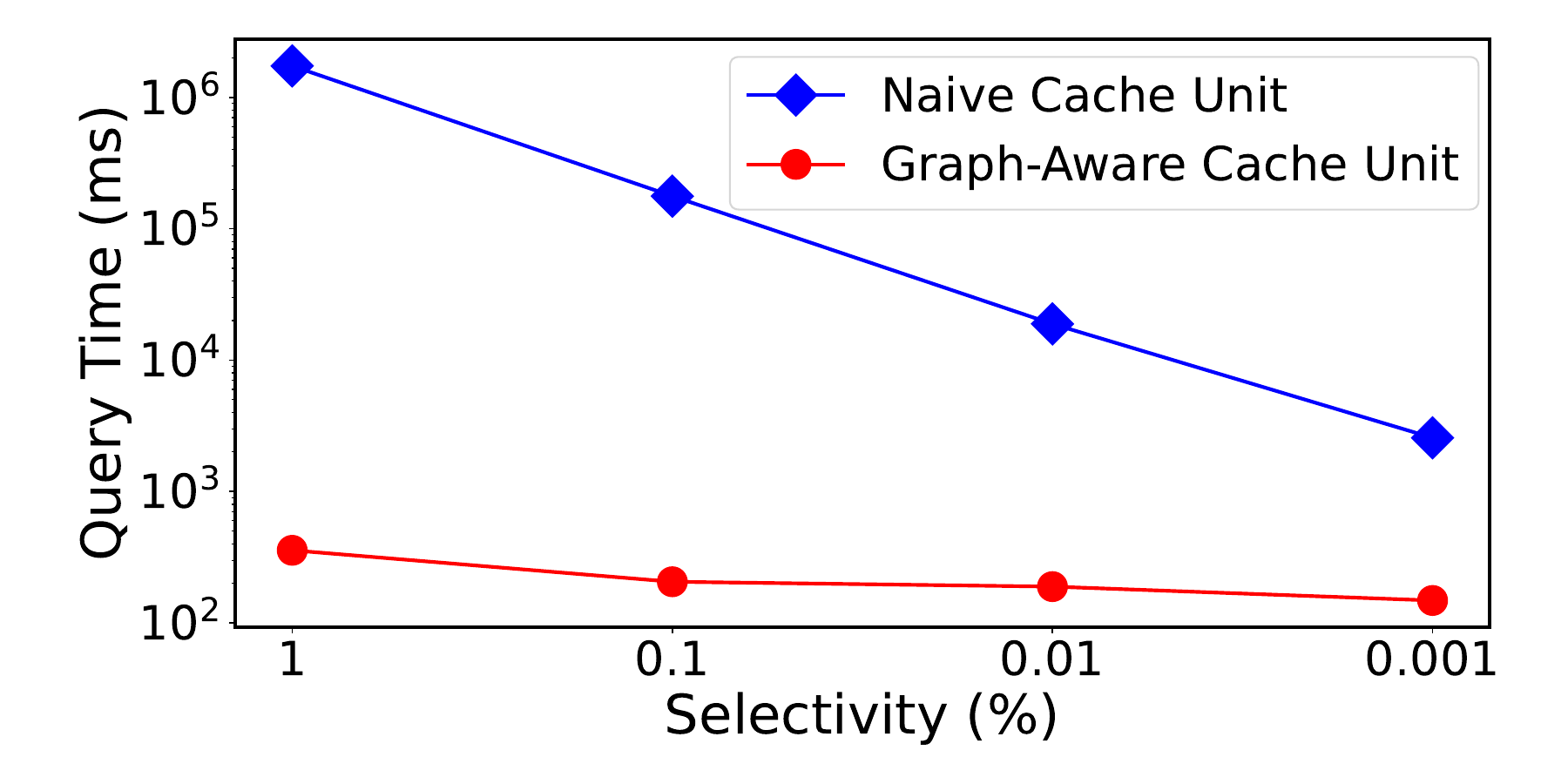}
\caption{Graph-aware cache units vs. naive column chunks}
\label{fig:graph_aware_cache_unit_study}
\end{figure}

\section{Related Work}\label{sec:relatedwork}


\myline{Relational Analytics over Lakehouse} 
With the trend toward Lakehouse architectures, prior work has focused on relational analytics over Lakehouse. For example, Databricks provides native SQL analytics that deliver state-of-the-art SQL performance~\cite{deltaengine2020}. Many relational engines also treat Lakehouse formats as first-class table sources~\cite{trino, Presto23, dremio, athena}. Several SQL warehouses are extending their capabilities to support relational analytics over Lakehouse~\cite{LevandoskiCDDEH24, DuckLake25, RedshiftLakehouse, SnowflakeLakehouse, SingleStoreLakehouse24}. However, these works focus on relational analytics over Lakehouse, while our work focuses on graph analytics over Lakehouse, prioritizing the graph topology and the structural dependencies between edges and vertices.

\myline{Graph Analytics in Relational Databases}
There is another line of work on evaluating graph queries over relational databases~\cite{hassan2018grfusion,jin2021making,graphprojection, duckpgq}. These systems typically build CSR-style graph structures to accelerate graph traversals, which is effective once the CSR is materialized. However, they incur non-trivial construction costs and target a different query-processing model than \sys{}, which must start directly from Lakehouse tables. In addition, \cite{graphprojection} proposes caching techniques for graph queries, but assumes access to tuple-addressable relational tables where properties can be fetched by row ID. Attributes in Lakehouse tables, however, are compressed and stored as column chunks in object storage, so their caching scheme cannot be directly applied to external columnar files. Spanner Graph~\cite{SpannerGraph24} uses interleaving to collocate edge rows with their source vertex rows, but this technique cannot be applied to our scenario since we cannot control the Lakehouse tables.

It is worth noting that GRFusion~\cite{hassan2018grfusion} also decouples graph topology as an index from graph element properties. However, it is designed for an in-memory setting rather than a Lakehouse setting. In addition, it lacks optimization for fast startup loading, in contrast to our work.


\myline{Disk-based Graph Databases} 
Our setting is conceptually related to disk-based graph databases~\cite{graphchi,lsmgraph,railwaygraph,FlashGraph}, since we can view the object store (and Lakehouse) as a disk layer. However, these systems are designed around fixed-size blocks and page-based buffer managers, whereas \sys{} operates on variable-size objects (e.g., column chunks and files) in remote storage and must optimize for fast startup loading and efficient graph analytics. In addition, prior disk-based graph engines typically employ CSR or CSR-like topology representations that associate each vertex with all of its neighbors. In contrast, \sys{} uses edge lists, which are explicitly optimized for fast construction from Lakehouse columnar data and potential updates over existing tables.

\myline{Others} Apache Spark~\cite{spark} is a prominent analytical engine for Lakehouse. Though its GraphX~\cite{gonzalez2014graphx} extension enables graph algorithms via RDD-based abstractions, it lacks native support for the labeled property graph model, which is a primary requirement for our system.

\section{Conclusion}\label{sec:conclusion}

In this work, we present \sys{}, a purpose-built graph compute engine built on top of TigerGraph to support graph analytics over Lakehouse tables. \sys{} introduces a suite of novel techniques to reduce startup time while ensuring query performance. Extensive experiments show that \sys{} significantly outperforms PuppyGraph, with substantially lower startup time and query time.



\bibliographystyle{ACM-Reference-Format}
\bibliography{paper}

\end{document}